\documentclass[sigconf]{acmart}
\AtBeginDocument{%
  }

\setcopyright{acmlicensed}
\copyrightyear{2026}
\acmYear{2026}
\acmDOI{XXXXXXX.XXXXXXX}
\acmConference[ E-ENERGY ’26]{The ACM International Conference on Future and Sustainable Energy Systems}{June 23–-25, 2026}{Banff, Canada }
\acmISBN{978-1-4503-XXXX-X/2026/06}

\usepackage{subfigure}
\usepackage{tikz}
\usepackage{subcaption}
\usepackage{pgfplots}
\usepackage{textcomp}
\usepackage{array}

\usepackage{amsmath,amssymb,amsfonts,mathtools}
\usepackage{algorithmic,tcolorbox}
\usepackage{algorithm}
\usepackage{graphicx}
\usepackage{bbding}
\usepackage{textcomp}
\usepackage{threeparttable}
\usepackage{framed}
\usepackage{verbatim}
\usepackage{bbm}
\usepackage{setspace}
\usepackage{mathrsfs}
\usepackage{bm}
\usepackage{url}
\usepackage{multirow}
\usepackage{verbatim}
\usepackage{float}
\allowdisplaybreaks[4]
\usepackage{hyperref}
\newcommand{\oh}[1]{#1}

\newcommand{\boxZ}{\ensuremath{Z}}
\newcommand{\boxZred}{\ensuremath{\tilde{Z}}}
\newcommand{\boxZdiam}{\ensuremath{C_Z}}
\newcommand{\cL}{\ensuremath{c_{\lambda}}}
\newcommand{\cLmax}{\ensuremath{\gamma_{\lambda}^{\max}}}
\newcommand{\A}{\ensuremath{A}}
\newcommand{\At}{\ensuremath{{\A^T}}}
\newcommand{\thep}{\ensuremath{\bar{p}}}
\newcommand{\thez}{\ensuremath{\bar{z}}}
\newcommand{\zOP}{\ensuremath{z^{\mathrm{op}}}}
\DeclareMathOperator{\INT}{\ensuremath{int}}

\newcommand{\SFF}[1]{\ensuremath{\sigma_{#1}}}
\newcommand{\INDF}[1]{\ensuremath{\delta_{#1}}}
\DeclareMathOperator{\dom}{\ensuremath{dom}}

\def\blue{\color{black}}

\begin{document}

\title{Can Carbon-Aware Electric Load Shifting Reduce Emissions?
An Equilibrium-Based Analysis}

\author{Wenqian Jiang}
\email{wenqian.jiang@wisc.edu}
\affiliation{%
  \institution{University of Wisconsin-Madison, Department of Electrical and Computer Engineering}
  \city{Madison}
  \state{Wisconsin}
  \country{USA}
}


\author{Olivier Huber, Michael C. Ferris}
\email{ohuber2@wisc.edu, ferris@cs.wisc.edu}
\affiliation{%
  \institution{University of Wisconsin–Madison, Department of Computer Sciences}
  \city{Madison}
  \state{Wisconsin}
  \country{USA}}

\author{Line Roald}
\email{roald@wisc.edu}
\affiliation{%
  \institution{University of Wisconsin-Madison, Department of Electrical and Computer Engineering}
  \city{Madison}
  \state{Wisconsin}
  \country{USA}
}






\begin{abstract}
An increasing number of electric loads, such as hydrogen producers or data centers, can be characterized as carbon-sensitive, meaning that they are willing to adapt the timing and/or location of their electricity usage in order to minimize carbon footprints. However, the emission reduction efforts of these carbon-sensitive loads rely on carbon intensity information such as average carbon emissions, and it is unclear whether load shifting based on these signals effectively reduces carbon emissions. {\blue To address this open question, we design a carbon-aware equilibrium model, which expands the commonly used equilibrium model for standard (carbon-agnostic) electricity market clearing to include carbon-sensitive consumers that adapt their consumption based on average carbon emission signals and carbon costs. This analysis represents an idealized situation for carbon-sensitive consumers, where their carbon preferences are reflected directly in the market clearing, and contrasts with current practice, where carbon emission signals only become known to consumers a posteriori (i.e., after the market has already been cleared). Furthermore, we extend our model to consider temporal load shifting and time-varying maximum renewable generations. We employ illustrative three-bus examples and numerical simulations on the IEEE RTS-GMLC system to reveal the limitations of the widely adopted average carbon emission signal for guiding carbon emission reduction. Our model offers a novel perspective for evaluating the effectiveness of different carbon signals and contributes to new carbon signal design.
} 





\end{abstract}





\maketitle

\section{Introduction}
The Paris Agreement recognizes emissions abatement as a critical strategy for mitigating global warming. Due to both the significant emissions from electricity generation and the potential for reducing these emissions by increasing renewable and low-carbon generation technologies, the electricity sector has been at the forefront of implementing decarbonization policies. However, the inherent variability of renewable generation, which is highly dependent on weather conditions (e.g., wind speed, solar radiation), necessitates aligning electricity demand with supply availability to maintain reliability. 
Demand-side flexibility is a crucial component of this alignment, i.e., consumers must be incentivized to adjust their consumption in response to renewable generation surpluses or deficits. 
At the same time, a growing number of carbon-sensitive loads, such as data centers and hydrogen production facilities, are emerging. These entities actively optimize the timing and location of their electricity usage to reduce carbon footprints, driven either by pledges to environmental responsibility or economic incentives such as production tax credits. However, many of these consumers are unwilling to give up control of their power consumption, e.g., by participating in a market, and instead prefer to locally optimize their consumption based on externally provided information such as carbon intensity signals. In this paper, we try to explore the coordination of carbon-sensitive loads with the rest of the electricity market clearing through a novel equilibrium formulation for an electricity market clearing where we model the impact of average carbon emissions on consumers' decision-making.

\subsection{Related Works}

To mitigate carbon emissions in power systems, researchers and policy-
makers have considered cap-and-trade emissions trading schemes \cite{EUETs, CCTp, Etssummary} and carbon tax schemes \cite{murray2015british, CCt}. Emissions trading provides greater environmental certainty in controlling overall emissions compared to a carbon tax, which defines a fixed emission price without restricting the quantity of GHGs emitted over a certain period \cite{Etssummary, stavins2019carbon, goulder2013carbon}. Many differences between these two schemes fade with specific implementation decisions \cite{stavins2019carbon}. Generally, these mechanisms reduce system-wide emissions in power systems by mainly increasing the cost of carbon-intense generation, thus dispatching more low-carbon generation. 
However, the increased generation costs are ultimately paid by consumers via elevated electricity prices \cite{nazifi2021carbon}. An analysis of 
the European emissions trading system showed that emission reductions primarily arise from reductions in demand due to higher electricity prices, rather than increased dispatch of clean generation \cite{chen2008implications, sijm2005co2}. Further, it is worth noting that a uniform carbon tax or the cap-and-trade emission trading scheme cannot account for differences in individual consumers' preference for low-carbon power.

Whereas calculating and potentially penalizing carbon emissions from generation is a relatively straightforward task, 
the question of how to allocate emissions from a group of generators across a group of consumers is less obvious \cite{bushnell2008design} and subject to ongoing debate \cite{standard2014ghg}. 
A common approach is to define a carbon intensity signal (with the most commonly used and widely available signal being \emph{average carbon emissions} \cite{standard2014ghg}) which is computed based on the average rate of emissions per megawatt-hour (MWh) of electricity consumed \cite{bettle2006interactions,elemaps}. To differentiate the impacts of locations on the carbon reduction, some researchers have advocated for the use of \emph{locational marginal carbon emissions} \cite{hawkes2014long, siler2012marginal, goldsworthy2023use, lindberg2022using, gorka2025electricityemissions}, which provides information about how a change in the electricity consumption at a given location would impact total emissions. 
Other, more recently developed metrics include \emph{locational average carbon emissions} \cite{chen2024contributions}, which is based on the concept of carbon flow \cite{kang2015carbon, chen2024carbon}, and \emph{adjusted locational marginal emissions} \cite{gorka2025electricityemissions}, which combines characteristics of locational and average emissions.

While carbon intensity signals are generally used for carbon accounting, e.g., as outlined in \cite{standard2014ghg}, carbon-sensitive loads also leverage these signals to decide where and/or when to consume electricity.
In practice, such adjustment would happen in a sequential manner, where the electricity market is first cleared assuming a given load profile and carbon intensity signals are then calculated based on the market outcome. 
Subsequently, carbon-sensitive consumers utilize these carbon intensity signals to adapt their consumption across time and space, with the goal of achieving carbon reductions. The impact of these changes has been studied in the context of real-time market clearing \cite{lindberg2022using, lindberg2021guide}, where the updated load profile becomes the input to the next market clearing, and in the context of day-ahead market clearing \cite{gorka2025electricityemissions}, where the updated load profile is taken into account in the intra-day market. 
As observed in \cite{lindberg2022using, lindberg2021guide, gorka2025electricityemissions}, this sequential approach leads to suboptimal outcomes because consumers act on carbon intensity signals arising from the previous market clearing, without insight into how their (and other consumers/generators) actions may change the generation dispatch and potentially impact the carbon intensity signals in the next market clearing. 

Our recent work in \cite{jiang2025greening} 
develops a novel electricity market clearing model
that incorporates the allocation of emissions from generations to
loads and allows for consideration of consumer-side carbon costs. Continuing to further this line of research, in this paper, we design a game-theoretic model where all market participants jointly identify a solution to the market clearing by solving an equilibrium problem that includes average carbon signal and their impacts on electricity consumption. As we mentioned before, there are several different methods to define the carbon emission intensity of electricity consumption.
We choose to use average carbon signals because this metric is widely available, both from system operators and companies \cite{CAISO, tomo}, and because it is commonly used for carbon accounting under the Greenhouse Gas Protocol \cite{GCP, standard2014ghg}. 
\subsection{Contributions}
The major contributions of this paper are threefold. 
\begin{itemize}
    \item We develop a novel equilibrium model considering the average carbon signal, which assumes that consumers have “perfect" knowledge of their own carbon emission intensity. This formulation provides a new perspective to evaluate the effectiveness of different carbon signals and contributes to new carbon signal design. To the best of our knowledge, this is the first paper that investigates the impact of carbon-sensitive consumers on carbon reductions from a game-theoretic perspective. 
    \item We extend our equilibrium model to consider time-varying maximum renewable power availability and temporal load shifting, which enables the analysis of how average carbon signals impact temporal load shifting results when the total demand for each consumer is fixed across all time periods. 
    \item We demonstrate how carbon-sensitive consumers behave in the equilibrium model against benchmark models based on the IEEE RTS-GMLC system. Numerical results show that the average carbon intensity signal may not be the “right" signal for load shifting to achieve effective carbon reductions.
\end{itemize}



The remainder of the paper is organized as follows: Section \ref{sec2} describes our equilibrium model. 
Section \ref{sec3} extends our model by considering time-varying maximum renewable generation and temporal load shifting. 
In Section \ref{ns}, we show results and conclusions from our numerical study, while Section \ref{sec6} summarizes and concludes.

\section{Carbon-aware Equilibrium Electricity Market Clearing Model}
\label{sec2}




In this section, 
we introduce the equilibrium problem formulation.
After that, we employ a simplified three-bus system to illustrate how loads adapting to carbon signals impact the generation dispatch and carbon emissions of the overall power system. 

However, before diving into the details of our formulation, we first introduce some basic background on equilibrium models in the electric sector. Equilibrium models, or game-theoretic models, are commonly employed to investigate strategic behavior in deregulated electricity markets \cite{ventosa2005electricity,hobbs2007nash}. In such models, there are typically three (in the electricity pool model, which ignores transmission constraints) or four (in the transmission-constrained model) types of participants, namely generators, consumers, transmission owners, and ISO. In the traditional carbon-agnostic electricity market, generators, consumers, and transmission owners aim at maximizing their profits given electricity prices as input parameters, and the ISO determines the market clearing price given power generation and electricity consumption as input parameters. All participants are, between them, playing a noncooperative game, and thus the optimal solution to this game is defined as a Nash equilibrium, corresponding to a situation where no participant can improve their outcomes by unilaterally changing their decisions. Under a price-taking assumption (no strategic bidding or market power), the Nash equilibrium problem is equivalent to a standard market-clearing optimization problem, i.e., solutions from the equilibrium problem are the same as those from the optimization problem. \oh{The equilibrium can often be rewritten as a complementarity problem; more details, examples and formulations can be found in } \cite{ferris1997engineering,gabriel2012complementarity, ferris2025optimizing}.




%

\subsection{Model Formulation}
\label{sub-21}

We next introduce our equilibrium formulation with carbon-sensitive loads.
As in the typical equilibrium model for electricity markets, we consider generators, consumers, transmission owners, and the ISO as participants. 
As is common, generators determine their optimal power generation based on electricity prices and the transmission owner maximizes their profit by buying power at one bus and selling it back at another one. Different from the standard model, consumers determine their electricity consumption considering \emph{both electricity prices and carbon emissions}, and the ISO determines both the price and the average carbon signal. Below, we provide the parametric optimization problems solved by the generators, consumers, and transmission owners, respectively, and \oh{the complementarity conditions involving the} ISO.





We consider an electric power network with the set of nodes, loads, transmission lines, and generators denoted by $\mathcal{N}$, $\mathcal{D}$, $\mathcal{L}$ and $\mathcal{G}$, respectively. Let $\mathcal{G}_i\subset \mathcal{G}$ and $\mathcal{D}_i\subset \mathcal{D}$ be the subset of generators and loads connected to the node $i$, and $(i,j)\in\mathcal{L}$ denote the transmission line from node $i\in \mathcal{N}$ to node $j\in \mathcal{N}$. Let $g\in \mathcal{G}$ and $d\in \mathcal{D}$ denote the indexes for generators and consumers, respectively.

\textbf{Generators}: Similar to the conventional electricity market, each generator in our model aims to maximize profit under generation constraints. Specifically, given the electricity price $p_{i}$ as input, the generator solves the following optimization problem to determine the optimal generation output that maximizes profits,
\begin{subequations}
\label{eq22}
    \begin{align}
    \max_{P_{G,g}} \ &(p_{i:g\in \mathcal{G}_i}-c_{G,g})\cdot P_{G,g}\label{eq22obj}\\
    s.t. \ &P_{G,g}^{\min}\leq P_{G,g}\leq P_{G,g}^{\max},\label{eq22a}
\end{align}
\end{subequations}
where $P_{G,g}$ is the optimal generation output (decision variables), $c_{G,g}$ is the generation cost for generator $g$, $p_{i:g\in \mathcal{G}_i}$ represents that the electricity price at the node $i$ to which the generator $g$ is connected, and the constraint (\ref{eq22a}) ensures that the chosen generation is feasible with $P_{G,g}^{\min}$ and $P_{G,g}^{\max}$ being generation limits.
    
\textbf{Consumers}: Each consumer in our model maximizes their benefits of consuming electricity, but different from a standard formulation, the consumers consider both price and carbon emissions in their decisions. This gives rise to a \emph{carbon-dependent} consumer utility, which we define as $r_{D,d}-\lambda\cdot c_{D,d}$. Here, $r_{D,d}$ is the revenue for consumers derived from consuming electricity, $c_{D,d}$ denotes the carbon cost, given in units of [\$/tons $CO_2$] and $\lambda$ is the value of the average carbon signal.
Specifically, given the price $p_i$ and the average carbon signal $\lambda$ calculated by ISO in the latter part, each consumer solves the following problem:
\begin{subequations}
\label{eq33}
\begin{align}
    \max_{P_{D,d}} \ &(r_{D,d}-p_{i:d\in \mathcal{D}_i}-\lambda\cdot c_{D,d})\cdot P_{D,d}\label{eq33obj}\\
    s.t. \ &P_{D,d}^{\min}\leq P_{D,d}\leq P_{D,d}^{\max}, \label{eq33a}
\end{align}
\end{subequations}
where $p_{i:d\in \mathcal{D}_i}$ represents the electricity price at the node $i$ to which the consumer $d$ is connected.
Each consumer incorporates the carbon term $\lambda \cdot c_{D,d}$ into their utility evaluation. {\blue In practice, the cost parameter $c_{D,d}$ may reflect concrete financial commitments, such as carbon taxes or the purchase of renewable energy certificates. It can also represent an internally defined “carbon cost", reflecting a consumer’s willingness to forgo revenue to avoid carbon emissions. For example, if a consumer is willing to pay \$100/MWh for zero-carbon electricity but only \$50/MWh for electricity emitting 1 tCO$_2$/MWh, the implied carbon cost is \$50/tCO$_2$.} 
Note that besides using carbon-dependent utilities to quantify the impacts of carbon emissions on consumers' decision-making,  
there are also other ways to quantify this impact, e.g., setting carbon emission limits \cite{wang2021optimal, pietzcker2021tightening}. However, it can be very challenging for consumers to define what the values of the carbon emission cap should be.

\textbf{Transmission Owners}: Transmission owners act as spatial arbitragers and aim
to maximize the power transmission profit given electricity prices $p_i$.
Specifically, the transmission owner maximizes their profit by buying power at a bus
and selling it back at another one (arbitrage through the electricity price difference
between two bus nodes), which is characterized by the following mathematical optimization
problem:
\begin{subequations}
\label{eqtr}
    \begin{align}
    \max_{\theta_i} \ &\sum_{i\in \mathcal{N}}p_i\cdot \left(\sum_{j,(i,j)\in \mathcal{L}}\beta_{ij}(\theta_{j}-\theta_i)\right)\\
    s.t. \ &-F_{ij}^{\rm{lim}}\leq\beta_{ij}(\theta_i-\theta_j)\leq F_{ij}^{\rm{lim}}, \quad\forall (i,j)\in \mathcal{L},\label{eqtra}
\end{align}
\end{subequations}
where $\beta_{ij}\in \mathbb{R}$ denotes the susceptance value of the transmission line $(i,j)$ from node $i$ to node $j$.
Constraints (\ref{eqtra}) are the transmission line limits, where $F_{ij}^{\rm{lim}}$ represents the transmission capacity (which we assume is the same in both directions).
{\blue Note that in the above, the voltage angle at the reference node $\theta_\mathrm{ref}$ is not a variable but the constant $0$.}

\textbf{ISO}: The ISO clears the market by determining the electricity trading price at each node $i$
and provides the average carbon signal.   
{\blue \oh{Specifically, they set each nodal electricity price $p_i$ such that the supply-demand power balance
\begin{equation}
   \quad \sum_{d\in \mathcal{D}_i}P_{D,d}+\sum_{j:(i,j)\in \mathcal{L}}\beta_{ij}(\theta_i-\theta_j)=\sum_{g\in \mathcal{G}_i}P_{G,g}\ \qquad(\perp p_i),\label{eq77c}   
\end{equation}
holds.
The average carbon signal $\lambda$ is defined as the total carbon emission divided by total demand, which can be written as
\begin{equation}
 \qquad\qquad\qquad   \lambda\sum_{d \in \mathcal{D}} P_{D,d} = \sum_{g \in \mathcal{G}}e_{G,g}\cdot P_{G,g} \qquad\qquad  (\perp \lambda), \label{eq923}
\end{equation}
where $e_G=\{e_{G,g}: \forall g \in \mathcal{G}\}$ is the carbon emission intensity vector for all generators (for generators using renewable sources, we assume that the carbon emission intensity values are zero).

Note that \eqref{eq77c} and \eqref{eq923} are written here as equality constraints but form a \emph{square} complementarity system when paired with $p$ and $\lambda$, respectively.
The number of equations in~\eqref{eq77c} (resp.~\eqref{eq923}) matches the size of the variable $p$ (resp.~$\lambda$).
Both $p$ and $\lambda$ are parameters in the optimization problems~\eqref{eq22}--\eqref{eqtr}.} 




The overall Nash equilibrium problem consists of the collection of parametric optimization problems \eqref{eq22}, \eqref{eq33}, \eqref{eqtr} and the parametric complementarity conditions \eqref{eq77c} and \eqref{eq923}. 
We note that there exist alternative, equivalent formulations of this  Nash equilibrium problem, which can be obtained by grouping optimality conditions differently. We show one such alternative formulation, which establishes a closer connection to the standard OPF market clearing, in Appendix \ref{appenef}.
}

{\blue We assume that there exists an interior feasible point (no variable $[P_{G,g},\forall g\in \mathcal{G}]$, $[P_{D,d},\forall d\in \mathcal{D}]$, and $[\theta_i,\forall i\in \mathcal{N}\setminus{\rm ref}]$ are at their bounds) such that the constraints ~\eqref{eq77c} and $\theta_{\rm ref}=0$ are satisfied.
This ensures the existence of a solution to our equilibrium problem, see Appendix~\ref{appendiex}.}

When all $c_{D}=0$, our equilibrium problem reduces to the standard social welfare maximization problem. The equilibrium framework seeks to coordinate all participants toward a consistent set of values for $P_G$, $P_D$, $p$, and $\lambda$ that simultaneously ensures power balance and aligns with the optimal solutions of each participant’s individual optimization problem. 

Unlike the sequential method, where inaccuracies may arise from relying on lagged market-clearing signals (see \cite{gorka2025electricityemissions, lindberg2022using, lindberg2021guide}), the equilibrium method uses synchronized decision-making based on perfect real-time carbon signals (as demonstrated by the alternative formulation in Appendix A), which we believe is  better for analyzing the effectiveness of different carbon signals. {\blue We note that, similar to the standard (carbon-agnostic) market-clearing optimization problem, the equilibrium problem may have multiple solutions (e.g., when multiple generators at a node share identical costs).}

\subsection{Three-bus Example Illustration}
In order to illustrate the outcomes of using an equilibrium model with carbon-dependent utility for consumers, we employ a simple three-bus system adapted from Example 6.2.2 in \cite{gabriel2012complementarity}, as seen in Fig. \ref{fig5}. The network 
includes three generators and three consumers, with one generator and consumer located at each bus. Since the carbon intensity of the generators is not included in the original three-bus system data, we define two carbon emission intensity cases
to see different results of load shifting among consumers in Table \ref{tab2}. In Case I, we assume that higher-cost generators have lower carbon emissions. Specifically, the most expensive generator, located at Bus 2, has the lowest carbon intensity, while the cheapest generator, located at Bus 3, has the highest carbon intensity. The medium-cost generator at Bus 1 has an intermediate carbon intensity. In Case II, we assume that higher-cost generators have higher carbon emissions. We assume that all consumers have a carbon cost of $c_{D,d} = \$20/$tons.

\begin{table}[hb]
\footnotesize
\centering
\caption{Parameters of Generators and Consumers.}
\label{tab2}
\begin{tabular}{ccccc}
\hline
&Bus (\#)& 1&2 &3\\
\hline
\multirow{3}{*}{Consumers}&$P_{D}^{\rm{min}}$(MW)&4&16&12\\
&$P_{D}^{\rm{max}}$(MW)&6&24&18\\
&$r_D$(\$/MWh)&18&20&21\\
\hline
\multirow{3}{*}{Generators}&$P_{g}^{\rm{min}}$(MW)&0&0&0\\
&$P_{G}^{\rm{max}}$(MW)&20&10&25\\
&$c_G$(\$/MWh)&8&10&6\\
Case I& \multirow{2}{*}{$e_G$(tCO$_2$/MWh)}&0.6&0.2&1\\
Case II&  &0.6&1&0.2\\
\hline
\end{tabular}
\end{table}





\begin{figure}
\centering
\subfigure[Case I. Higher cost generators have lower carbon intensity.]{
\includegraphics[width=0.82\linewidth]{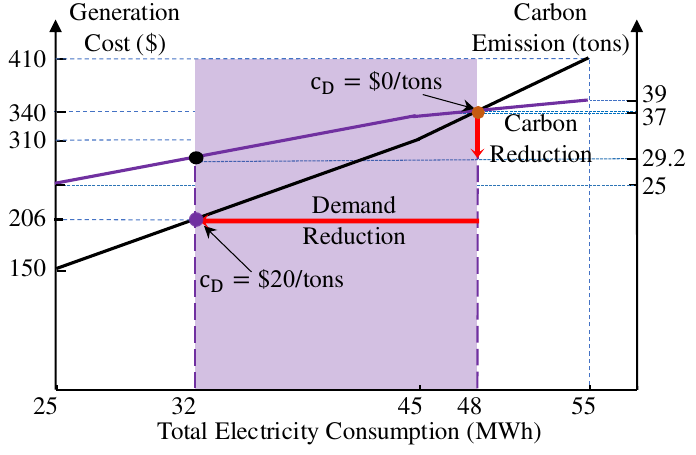}
}

\subfigure[Case II. Higher cost generators have higher carbon intensity.]{
\includegraphics[width=0.92\linewidth]{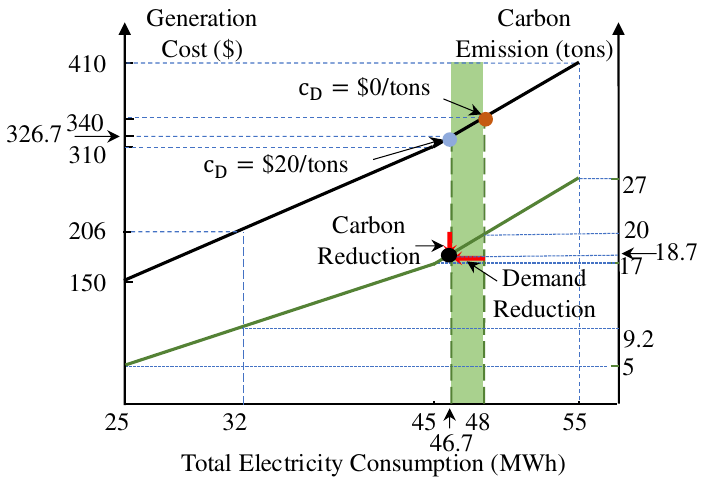}
}

\caption{Results for different generation cost-carbon intensity pairs.}
\label{fig4}
\end{figure}

\begin{figure}
\centering
\subfigure[Case I.]{
\includegraphics[width=0.44\linewidth]{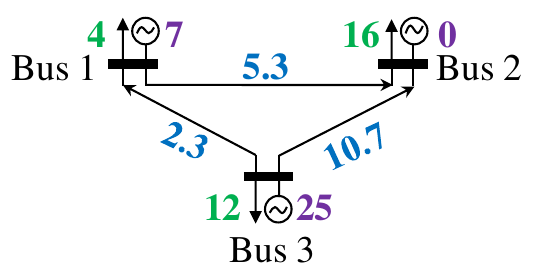}
}
\quad
\subfigure[Case II.]{
\includegraphics[width=0.42\linewidth]{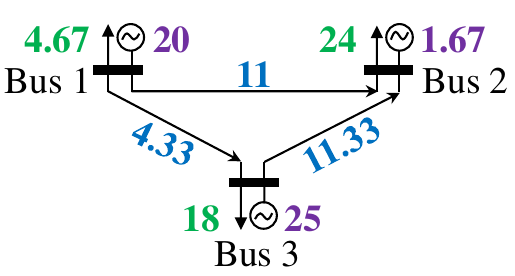}
}
\caption{Generation and consumption for Case I and Case II in Table \ref{tab2}.}
\label{fig5}
\end{figure}

\subsubsection{Electricity Pool Case} We first consider the electricity pool case, which omits consideration of transmission constraints. The results for this case are shown in Fig. \ref{fig4} (a) and (b) and detailed generation dispatch (purple numbers) and demand values (green numbers) are shown in Fig. \ref{fig5}. Fig. \ref{fig4} (a) shows results for Case I, with the black line showing the generation cost curve (with generators ordered from the cheapest to the most expensive) and the purple line showing the total carbon emission curve for Case I. The total carbon emissions $E_{tot}$ is calculated by multiplying the amount of generation with the corresponding carbon intensities, i.e., $E_{tot}=\sum_{g\in\mathcal{G}}e_{G,g}\cdot P_{G,g}$ based on the generation dispatch order. The red and purple points on the black generation cost curve represent the dispatch results for two cases, respectively. The purple shadowed areas highlight the effects of our model compared with the case without considering carbon emissions. 
Similarly, Fig. \ref{fig4} (b) shows results for Case II, with the black line showing the generation cost curve (which is the same as for Case I) and the green line showing the total emission curve for Case II.


Our first observation from Fig. \ref{fig4}, is that the only impact of considering the carbon-dependent utility is that the consumers are less willing to pay for electricity with a non-zero average carbon footprint and hence reduce their electricity demand. 
The generation merit order (represented by the generator cost curve) creates an ordering of the generators.
The demand reduction sequentially impacts the generators, starting with the marginal one, which reduces production, and if needed, continues in descending order.
If the affected generators have non-zero emissions, reductions in demand have the intended effect of decreasing emissions, as witnessed in both Case I and Case II. 
However, the fact that carbon-aware
loads can only sequentially impact the generators according to the aforementioned ordering, which is not related to the ranking of generators by their emissions, is a fundamental limit of carbon-aware load shifting in our equilibrium model.

Our second observation from Fig. \ref{fig4} is that the demand reduction and associated reduction in emissions is very different between Case I and II. In Case I, the generation dispatch includes low-cost, carbon-intensive generation which leads to a high average carbon value $\lambda=0.913$. Since $\lambda$ is high, the carbon-dependent consumer utility is much lower than the normal consumer utility, leading to a large reduction in demand from 48 to 32 MW, or 33\%, when considering carbon. However, since the marginal generator has low emissions, this reduction in demand only decreases emissions from 37 to 29.2 tons, or 21\%. This suggests that each unit of demand reduction achieves a lower-than-average impact on emissions in Case I. 
Conversely, in Case II, the generation dispatch includes low-cost, low-carbon generation which leads to a low average carbon value $\lambda=0.4$. With a lower $\lambda$, the carbon-dependent consumer utility is closer to the normal consumer utility, and the demand is only reduced to 46.7 MW, or less than 3\%, when considering carbon. This reduction in demand reduces emissions from 20 to 18.7 tons, or 6.5\%. Thus, in Case II, each unit of demand reduction achieves a higher-than-average impact on emissions, but the overall demand and carbon reductions are small. 

This example highlights why the average carbon signal may not be the best choice, as it does not provide any carbon intensity information about the marginal generator, which is the only generator impacted by the load shift, and thus does not accurately predict how a change in demand may increase or decrease emissions. Reducing demand in a situation with low average carbon emissions, such as Case II, can have a greater impact on emissions than reducing demand in situations with high average emissions, such as Case I. 

\begin{figure*}
\centering
\subfigure[Base case ($c_{D,d}=0$ for $\forall d\in \mathcal{D}$).]{
\includegraphics[width=0.3\linewidth]{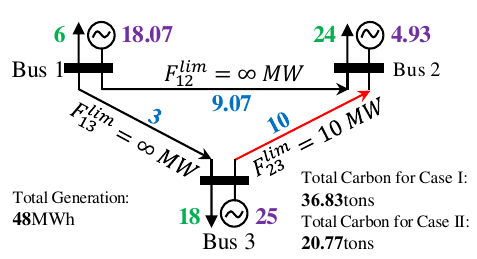}
}
\quad
\subfigure[Case I. Higher cost-lower carbon generators.]{
\includegraphics[width=0.3\linewidth]{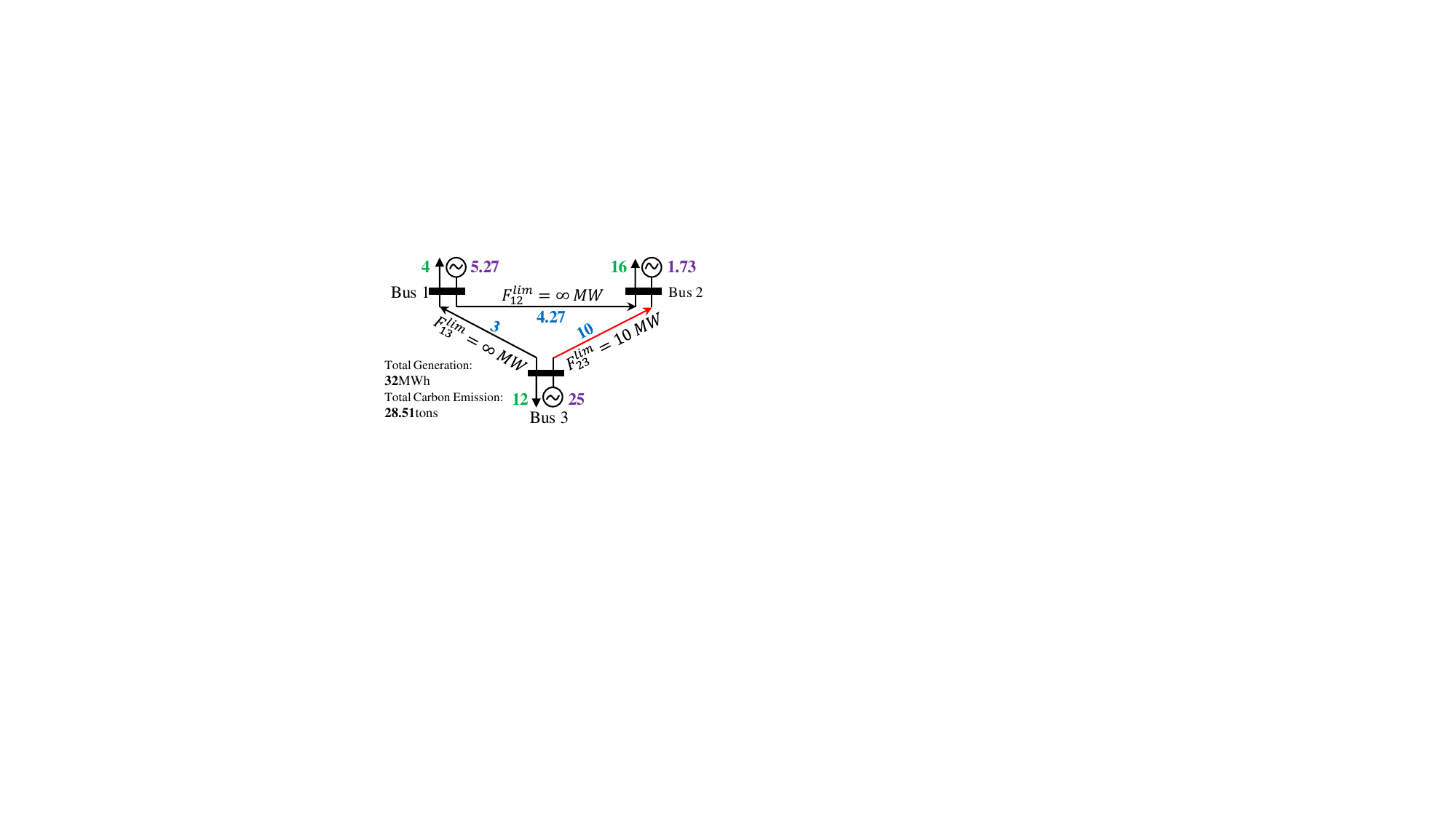}
}
\quad
\subfigure[Case II. Higher cost-higher carbon generators.]{
\includegraphics[width=0.3\linewidth]{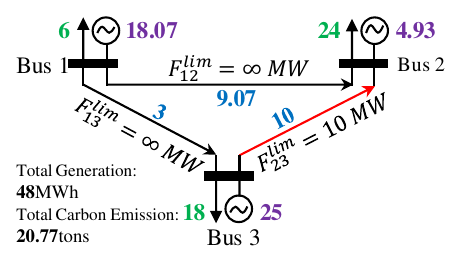}
}
\caption{Results comparison for transmission-constrained cases.}
\label{fig6}
\end{figure*}

\subsubsection{Transmission-constrained Case}
To explain the impacts of transmission line constraints in our equilibrium model,
we reduce the transmission capacity on the transmission line from Bus $2$ to Bus $3$ so it becomes congested. This congestion increases power generation at Bus 2 and decreases power generation at Bus 1.
Fig. \ref{fig6} shows the dispatch for the base case ($c_D=0$) and Case I and Case II ($c_D=20$), all with transmission line limits. 
For Case I, the shift of generation from generator 1 to 2 increases the average carbon emissions from $\lambda = 0.608$ (for the electricity pool case) to $\lambda=0.767$ in the base transmission-constrained case (Case I in Fig. 3a). Introducing carbon-sensitive loads reduces the load from 48 to 32 MW, as in the case without transmission constraints. However, due to congestion, both generators 1 and 2 are now marginal and are impacted by the demand reduction, leading to a slightly higher reduction in emissions of 22.5\%. Nevertheless, the average emissions increase to $\lambda=0.89$ after the load reduction.
For Case II, shifting generation from generator 1 to 2 reduces the average carbon emissions to $\lambda=0.433$. This $\lambda$ is low enough to allow consumers to continue to consume their maximum load, and as a result, the generation dispatch does not change after we introduce carbon-sensitive loads. This is despite the fact that the two marginal generators in this case have higher emissions than in Case I.  

This example highlights that transmission constraints impact the marginal generators serving carbon-sensitive loads and thus impact the effectiveness of demand reductions from carbon-sensitive loads.

\section{A Multi-Period Case with Temporal Load Shifting}
\label{sec3}
Section \ref{sec2} introduces an equilibrium model for a single time slot, in which consumers contribute to carbon reductions by curtailing their electricity demand from the maximum down to the minimum level when the carbon-dependent term $r_{D,d}-\lambda\cdot c_{D,d}$ is sufficiently small (specifically, when it falls below $p_{i:d\in \mathcal{D}_i}$). However, a more impactful and practical strategy for carbon reductions involves temporal load shifting, where consumers reallocate their fixed total electricity consumption across periods. To realistically model this strategy, it is crucial to also account for the time-varying nature of renewable generation. Thus, in this section, we formulate the equilibrium model in a multi-time-scale setting that considers both temporal load shifting and the dynamics of maximum renewable generation limits.



\subsection{Model Formulation}
Different from the single-period model in Section \ref{sec2}, each participant in the multi-period model needs to optimize their behaviors across considered periods under time-varying parameters, and detailed optimization problems are introduced below.

\textbf{Generators}: Given the electricity price at each time $p_{i,t}$ as input, the generator solves the following optimization problem to maximize their profit across $T$ time periods:
\begin{subequations}
\label{eq22g}
 \begin{align}
    \max_{P_{G,g,t}} \ &\sum_{t\in \mathcal{T}}(p_{i:g\in \mathcal{G}_i,t}-c_{G,g,t})P_{G,g,t}\\
    s.t. \ &P_{G,g}^{\min}\leq P_{G,g,t}\leq  P_{G,g,t}^{\max}, \ \forall t,\label{eq22ga}
\end{align}   
\end{subequations}
where $\mathcal{T}=[1,2,\cdots,T]$.
We assume that the maximum power $P_{G,g,t}^{\max}$ of all renewable generators are time-varying values, whereas the non-renewable values are fixed across time.


\textbf{Consumers}: Given time-varying electricity prices $p_{i,t}$ and average carbon signals $\lambda_t$ calculated by ISO in the later part, each consumer maximizes their utilities by solving the following problem: 
\begin{subequations}
    \label{age}
     \begin{align}
      \max_{P_{D,d,t}} \ &\sum_{t\in \mathcal{T}}(r_{D,d,t}-p_{i:d\in \mathcal{D}_i,t}-\lambda_{t}\cdot c_{D,d,t})\cdot P_{D,d,t} \\
      s.t. \ & P_{D,d}^{\min}\le P_{D,d,t}\le P_{D,d}^{\max},\ \forall t,\label{agea}\\
      &\sum_{t\in \mathcal{T}}P_{D,d,t}=\mathcal{P}_{D,d}.\label{agec}
\end{align}
\end{subequations}
where $\mathcal{P}_{D,d}$ is the fixed total electricity consumption across $T$ time slots for the consumer $d$. Note that in the multi-period case, the load schedule is dictated by the priority order of $r_{D,d,t} - p_{i:d \in \mathcal{D}_i,t} - \lambda_{t} \cdot c_{D,d,t}$ across different time periods $t$. Specifically, consumption is prioritized for time slots with the larger values of $r_{D,d,t} - p_{i:d \in \mathcal{D}_i,t} - \lambda_{t} \cdot c_{D,d,t}$.



\textbf{Transmission Owners}: The transmission owner aims to maximize their profits across $T$ time periods: 
\begin{subequations}
\label{eqtrt}
    \begin{align}
    \max_{\theta_{i,t}} \ &\sum_{t\in \mathcal{T}}\sum_{i\in \mathcal{N}}p_{i,t}\cdot \left(\sum_{j:(i,j)\in \mathcal{L}}\beta_{ij}(\theta_{j,t}-\theta_{i,t})\right)\\
    s.t. \ &-F_{ij}^{\rm{lim}}\leq\beta_{ij}(\theta_{i,t}-\theta_{j,t})\leq F_{ij}^{\rm{lim}}, \quad\forall (i,j)\in \mathcal{L}, \ \forall t\label{eqtra2},\\
    &\theta_{ref,t} = 0,\ \forall t\label{eqtb2}.
\end{align}
\end{subequations}

\textbf{ISO}: The ISO clears the market by determining electricity prices and provides average carbon signals. Specifically, they set electricity prices $p_{i,t}$ by ensuring the supply-demand balance constraint 
\begin{equation}
    \sum_{d\in \mathcal{D}_i}P_{D,d,t}+\sum_{j:(i,j)\in \mathcal{L}}\beta_{ij}(\theta_{i,t}-\theta_{j,t})=\sum_{g\in \mathcal{G}_i}P_{G,g,t}, \ \forall i\in \mathcal{N}, \ \forall t. \label{eq77m}
\end{equation}
Average carbon emissions can be calculated by:
\begin{equation}
    \lambda_{t}\sum_{d\in \mathcal{D}}P_{D,d,t} = \sum_{g \in \mathcal{G}}e_{G,g,t}\cdot P_{G,g,t},\ \forall t.  \label{eq923g2}
\end{equation}
Both $p_{i,t}$ and $\lambda_t$ are parameters in the optimization problems (\ref{eq22g})-(\ref{eqtrt}). 

Similar to the single-period model, the overall Nash equilibrium problem for the multi-period case consists of the collection of parametric optimization problems \eqref{eq22g}, \eqref{age}, \eqref{eqtrt} and the parametric complementarity conditions \eqref{eq77m} and \eqref{eq923g2}.






\subsection{Three-bus Example Illustration}
\label{sub33}
To understand the impacts of average carbon signals on our multi-period equilibrium model with temporal load shifting, we consider the transmission-constrained three-bus example in Section \ref{sec2}. Specifically, we consider Case I in Table \ref{tab2}, under which the generator $2$ owns the highest generation cost but the lowest carbon factor of generating unit power, of which we assume the maximum power limit is time-varying. To make it simple, we only consider $T=3$ (longer time periods and more complex cases will be discussed for the IEEE RTS-GMLC system in numerical studies), $r_{D,d,t}$, $e_{G,g,t}$, and $c_{G,g,t}$ are fixed for each consumer across time, and the maximum generation values for generators 1 and 3 are fixed for all time slots at $20$MW and $15$MW, respectively, whereas the maximum power limit for generator 2 varies according to $P_{G,2,t}^{\max}=[5,15,30]$MW. The transmission capacity on the transmission line from Bus $1$ to Bus $2$ is set at $F_{12}^{lim}=8$MW (with the remaining two transmission lines unlimited) and the total consumption for each consumer across all periods is $\mathcal{P}_{D,d} = 3*90\%*P_{D,d}^{\max}$.

\begin{table*}[t]
\footnotesize
\centering
\caption{\small Impact of Temporal Load Shifting on Power Dispatch, Load Dispatch, Carbon Emissions, and Generation Costs.}
\label{tab:temporal_results}
\begin{tabular}{>{\centering\arraybackslash}p{1.2cm}|
                >{\centering\arraybackslash}p{0.2cm} >{\centering\arraybackslash}p{0.2cm} >{\centering\arraybackslash}p{0.2cm}|
                >{\centering\arraybackslash}p{0.2cm} >{\centering\arraybackslash}p{0.2cm} >{\centering\arraybackslash}p{0.2cm}|
                >{\centering\arraybackslash}p{0.2cm} >{\centering\arraybackslash}p{0.2cm} >{\centering\arraybackslash}p{0.2cm}|
                >{\centering\arraybackslash}p{0.2cm} >{\centering\arraybackslash}p{0.2cm} >{\centering\arraybackslash}p{0.2cm}|
                >{\centering\arraybackslash}p{0.2cm} >{\centering\arraybackslash}p{0.2cm} >{\centering\arraybackslash}p{0.2cm}|
                >{\centering\arraybackslash}p{0.2cm} >{\centering\arraybackslash}p{0.2cm} >{\centering\arraybackslash}p{0.2cm}|
                >{\centering\arraybackslash}p{0.2cm} >{\centering\arraybackslash}p{0.2cm} >{\centering\arraybackslash}p{0.2cm}|
                >{\centering\arraybackslash}p{0.2cm} >{\centering\arraybackslash}p{0.2cm} >{\centering\arraybackslash}p{0.2cm}|
                >{\centering\arraybackslash}p{0.6cm} >{\centering\arraybackslash}p{0.6cm} >{\centering\arraybackslash}p{0.6cm}}
\hline
\multirow{2}{1.2cm}{\centering \textbf{Case definition}} & \multicolumn{9}{c|}{\textbf{Power dispatch}}& \multicolumn{9}{c|}{\textbf{Load dispatch}}
& \multicolumn{3}{c|}{\textbf{Ave. Carbon}} &\multicolumn{3}{c|}{\textbf{Tot. Carbon}}&\multirow{2}{1.3cm}{\textbf{Gen \\Cost}}&\multirow{2}{1.3cm}{\textbf{Tot. \\Carbon}} &\multirow{2}{1.3cm}{\textbf{Ave. \\Carbon}}\\
\cline{2-10} \cline{11-19} \cline{20-25}
& $g_{1,1}$ & $g_{1,2}$ & $g_{1,3}$ & $g_{2,1}$ & $g_{2,2}$ & $g_{2,3}$ & $g_{3,1}$ & $g_{3,2}$ & $g_{3,3}$ & $d_{1,1}$ & $d_{1,2}$ & $d_{1,3}$ & $d_{2,1}$ & $d_{2,2}$ &$d_{2,3}$ & $d_{3,1}$ & $d_{3,2}$ & $d_{3,3}$ &$\lambda_{1}$&$\lambda_2$&$\lambda_3$&$t_1$&$t_2$&$t_3$&&&\\
\hline
\textbf{$c_{D,d}=0$} &19.5&19.9&18.7&3.9&11.5&11.1&15&15&15&5.9&5.4&4.9&17.1&23.7&24&15.4&17.3&15.9 &0.72&0.63&0.64&27.6&29.2&28.7&999.6&85.2&0.66\\
\hline
\textbf{$c_{D,d}=30$} & 16.6&20&20&2&13&13&15&15&15&4.2&6&6&16.8&24&24&12.6&18&18&0.75&0.62&0.62&25.2&29.8&29.8&1002.9&84.6&0.65\\
\hline
\end{tabular}
\end{table*}
Table \ref{tab:temporal_results} presents a comparative analysis of temporal load shifting between the base case $c_D = 0$ and the case with $c_D = 30$. Focusing on the consumer’s problem in (\ref{age}), we have that given the same $r_{D,d,t}$ across all time, if the condition
\begin{align}
    p_{i:d\in \mathcal{D}_i,(1)}&+\lambda_{(1)}\cdot c_{D,d}\ge p_{i:d\in \mathcal{D}_i,(2)}+\lambda_{(2)}\cdot c_{D,d}\ge \notag\\ 
    &\cdots \ge p_{i:d\in \mathcal{D}_i,(T)}+\lambda_{(T)}\cdot c_{D,d}
\end{align}
holds, then the corresponding power consumption follows 
\begin{equation}
    P_{D,d,(1)}\le P_{D,d,(2)}\le \cdots \le P_{D,d,(T)}.
\end{equation}
This can be validated from the results in the case of $c_D = 30$. The prices $p_{i,t}$ are all 10\$/MWh except for $p_{1,1}=8$\$/MWh and $p_{3,1}=9.1$\$/MWh under this case. The values of $p_{i:d\in \mathcal{D}_i,t}+\lambda_{t}\cdot c_{D,d}$ at the latter two time slots are smaller than that at $t=1$. Hence, more electricity is consumed at the latter two time slots. Note that this load shifting is driven by lower carbon emissions at $t=2$ and $t=3$ (because of $p_{i,1}<p_{i,2}=p_{i,3}$ for consumers 1 and 3 and equal for consumer 2). This temporal load shifting aligns well with the trend of renewable power availability ($P_{G,2,t}^{\max}=[5,15,30]$).   


A comparison between $c_D = 0$ and $c_D = 30$ reveals that the influence of average carbon signals only marginally increases renewable generation (from Generator 2) from 26.5MWh to 28MWh. Consequently, total carbon emissions experience a slight decline from 85.2tCO$_2$ to 84.6tCO$_2$, accompanied by a cost increase from \$999.6 to \$1002.9. Another interesting observation is that although average carbon emissions decrease at $t=2$ and 3, total emissions at these times increase due to higher consumption levels happening at the same time for $c_D = 30$ (compared to $c_D=0$). Furthermore, in both cases, the generation dispatch results obtained by solving Problem (\ref{eq22g}) follow the generation cost merit order, with lower-cost generators being dispatched first. For instance, Generator 3, having the lowest cost, operates at its maximum output of 15 MWh across all time periods, while the total power output of Generator 1 consistently exceeds that of Generator 2, which has the highest cost.

This example highlights again why the average carbon signal may not be the best choice, as it promotes peak consumption occurring simultaneously, i.e., incentivizing increased consumption during periods of lower average carbon emissions. While this may reduce the average carbon intensity at certain times, it can lead to an overall increase in total carbon emissions during those periods (as shown in the above example). Furthermore, as noted in \cite{ruiz2010analysis}, depending on the specific power system conditions, certain consumers may need to take actions contrary to others in order to effectively reduce system-wide total carbon emissions. It is evident that average carbon signals lack the granularity needed to facilitate such coordinated and differentiated responses.

\section{Numerical Studies}
\label{ns}

 This section conducts extensive numerical studies to demonstrate how the equilibrium method could impact electricity market clearing. All the tests are performed on a laptop computer with an Apple M3 Pro CPU and 36GB RAM. {\blue Our equilibrium problem is reformulated as a system of constraints formed by the KKT conditions of Problems ~\eqref{eq22}~–~\eqref{eqtr}, together with the complementarity constraints ~\eqref{eq77c} and ~\eqref{eq923}. This system is then implemented in GAMS (or GAMSpy) \cite{GAMs} using the Extended Mathematical Programming (EMP) framework \cite{kim2019solving} and solved numerically using the PATH solver \cite{dirkse1995path}.} 
We adopt the IEEE RTS-GMLC system \cite{barrows2019ieee} for numerical analysis. This system has $73$ buses, $158$ generators, and $120$ lines. We assign emission factors for each generator based on their assigned fuel type and data from the US Department of Energy \cite{emissiondata}. Specifically, we assign $e_G=\{0.6042, 0.7434, 0.9606\}$ for natural gas, oil, and coal generators, respectively, and assume that solar, wind, and hydro have $e_G=0$. For simplicity, we consider each non-zero load in the system as a consumer, leading to a total of $51$ consumers located at different buses. We draw consumer utility values $u_D$ from a uniform distribution in the range of $[20,80]$~\$/MWh.
According to \cite{carbonp}, carbon prices in the worldwide emission trading markets range from $10$ to $100$\$/tons. Based on these statistics, we assume that consumers will have carbon prices $c_{D}$ in the range between 
$[0,100]$ in the following experiments. 
For consumers' demand flexibility, we use the values in the IEEE RTS-GMLC system as the maximum values and 80\% of these values as the minimum values. 



\subsection{Evaluation Metrics and Benchmark Formulations}
\label{sec41}
In our case study, 
we consider the following quantities when evaluating market-clearing solutions: 
\begin{itemize}
    \item Total generation: $\sum_{g\in \mathcal{G}}P_{G,g}$;
    \item Total generation cost: $\sum_{g\in \mathcal{G}}c_{G,g}P_{G,g}$;
    \item Total carbon emissions: $\sum_{g\in \mathcal{G}}e_{G,g}P_{G,g}$;
    \item Average carbon emissions: $\lambda$. 
\end{itemize}
We consider five benchmark models for our analysis: 
\begin{itemize}
    \item \textbf{Market Clearing with Fixed Demands (MCFD):} The traditional (carbon-agnostic) electricity market solves the market clearing problem assuming a fixed demand. For simplicity, we assume that the demand of each consumer is fixed at their maximum values $P_{D,d}^{\max}$.
    \item \textbf{Market Clearing with Demand Flexibility (MCDF):} In this case, consumers have flexibility in their electricity consumption but do not consider carbon costs. Note that this model is equivalent to a version of our equilibrium model where all consumers set carbon costs $c_{D}=0$.
    \item \textbf{Market Clearing with Minimizing Carbon Emissions (MCE):} For comparison, we also involve a case to minimize carbon emissions of the whole system, where constraints are the same as MCDF, but the objective function is to minimize the total carbon emission as follows:
    \begin{equation}
        \min_{P_G, P_D, \theta}\left(\sum_{g\in \mathcal{G}}e_{G,g}P_{G,g}\right).
    \end{equation}
    \item \textbf{Carbon Tax Method (CTM):} One common effort to reduce carbon emissions is adding carbon taxes on electric generators \cite{baranzini2000future,fischer2008environmental}. 
    Mathematically, the problem can be detailed as follows:
\begin{align}
\nonumber
    \max_{P_G, P_D, \theta}&\left(\sum_{d \in \mathcal{D}}r_{D,d} P_{D,d}-\sum_{g\in \mathcal{G}}\left(c_{G,g}+C_{co_2}e_{G,g}\right)P_{G,g}\right)\label{eqcon}\\
    s.t. \ & \rm{Constraints} \ (\ref{eq22a}), \ (\ref{eq33a}), \ (\ref{eqtra}),\ (\ref{eq77c})\notag.
\end{align}
where $C_{co_2}\in \mathbb{R}$ is the carbon tax, which is usually determined by energy-related governments. From a game-theoretic perspective, this carbon tax model is also equivalent to an equilibrium model, where each generator maximizes their profits considering both price and carbon emissions in their decisions. We defer the detailed equilibrium formulation for the carbon tax method to Appendix \ref{appen2}.
\item \textbf{Sequential Method (SM):} 
We consider the sequential method first developed to represent geographical load shifting in \cite{lindberg2020environmentalpotentialhyperscaledata} and extended to consider multiple time steps in  
\cite{gorka2025electricityemissions}. 
This method is representative of current practice and consists of three steps:

\underline{Step $1$:} We clear the day-ahead electricity market, without considering consumer carbon costs. This step is equivalent to solving the MCFD model.

\underline{Step $2$:} We calculate the average carbon emissions from the market clearing results, namely $\lambda_{b}$. We then solve \eqref{eq33} with the calculated average carbon emission to represent how carbon-sensitive loads would adapt their load.

\underline{Step $3$:} Finally, we solve another market clearing similar to the MCFD, representing the intra-day electricity market, with updated load profiles. We denote the average carbon emission based on these market clearing results as $\lambda_{a}$.

\qquad One critical issue with this sequential method is that there is a time lag in the signal $\lambda_e$ received by the consumer. Specifically, consumers only receive the carbon emission signal a posteriori, after the market is cleared, and do not know the impacts of their behaviors on the next market clearing results. 
This is in contrast to our equilibrium formulation, which allows consumers to possess “perfect" knowledge of carbon emission intensity signals. 

%
\end{itemize}

\subsection{Analysis for Different Carbon Costs}
\label{adcb}
We first analyze the impact of different carbon costs on the equilibrium results, assuming that all consumers are willing to submit non-zero carbon costs to help reduce carbon emissions. We compare cases where the carbon costs of consumers are randomly drawn from a uniform distribution with ranges $[10,40]$, $[30,60]$, $[50,80]$, and $[80,100]$, respectively. 
Table \ref{tab11} shows the results.
The top rows represent the two carbon-agnostic formulations (MCFD and MCDF), and the average carbon cost increases as we go further down in the rows.

\begin{table}[t]\footnotesize
\renewcommand{\arraystretch}{1.1}
\setlength{\tabcolsep}{3.2pt}
\centering
\caption{The impact of different carbon costs on generation dispatch, system emissions, and other components in equilibrium models.}
\label{tab11}
\begin{tabular}{ccccc}
\hline
Cases& \begin{tabular}[c]{@{}c@{}}Total Generation\\~[MWh]\end{tabular}  &\begin{tabular}[c]{@{}c@{}}Total Generation \\ Cost [\$]\end{tabular} &\begin{tabular}[c]{@{}c@{}}Total Carbon\\~[tons]\end{tabular}&\begin{tabular}[c]{@{}c@{}}Average Carbon\\~[tCO$_2$/MWh]\end{tabular}\\
\hline
MCFD&8550	&63748&	3001.8&	0.351\\
MCDF	&8550	&63748	&3001.8	&0.351	\\
$[10,40]$&8550&63748&3001.8	&0.351\\
$[30,60]$&8515.8	&63358&	2969&	0.349	\\
$[50,80]$&8368.4	&61677.3	&2827.4	&0.338\\
$[80,100]$&	8159.2	&59986.7	&2680.8	&0.329\\
\hline
\end{tabular}
\end{table}


We first observe that the two carbon-agnostic market clearing models  (MCFD and MCDF) lead to the same solutions. This indicates that the consumer utility values $r_D$ are high enough to cause loads to consume at their highest level in the formulation with load flexibility, but no carbon costs. Next, we consider the impact of low carbon costs in the range $[10,40]$ (the third row of Table \ref{tab11}). For these carbon costs, we observe a negligible impact on total power generation, indicating that the loads are still consuming the same amount of electricity despite occurring a carbon cost. We conclude that when carbon costs are low, they do not impact consumers' electricity consumption. Further, we consider the case with higher carbon costs in the ranges $[30,60]$, $[50,80]$, or $[80,100]$ (lower three rows of Table \ref{tab11}).
As the carbon costs increase to these levels, the total demand decreases, and the total generation decreases accordingly. This happens because some loads choose to consume less, as the carbon-dependent utility is too low (because the carbon cost becomes too high). As the total generation decreases, the total generation cost, total carbon emissions, and average carbon emissions also decrease.

{\blue \subsection{The Impacts of Carbon Costs on Electricity Prices and Generation Profits}
The introduction of carbon-dependent consumer utility will impact market clearing results, leading to different electricity prices faced by both generators and consumers as well as different generation profits. We next analyze how carbon-dependent utility, represented through carbon costs, impacts electricity prices and generation profits compared to those obtained from a standard market clearing (i.e., where all carbon costs are zero). To assess this, we assign consumer carbon costs $c_D$ in a range of $[30,60]$ \$/MWh, and then randomly assign zero carbon costs to 25\% of the consumers to simulate carbon-agnostic consumers. For consumers' demand flexibility, we use 120\% of values in the RTS-GMLC system as the maximums. 

In this scenario, the introduction of carbon costs causes consumers to use less electricity, and the total electricity generation decreases from 10,260 MWh in the standard benchmark to 10,093.6 MWh (-1.6\%).
The carbon emissions fall from 3,900.6 tCO$_2$ to 3,785.5 tCO$_2$ (-2.9\%), while the overall generation cost drops from \$77,910.4 to \$75,187.8 (-3.6\%). These results suggest that a small reduction in loads leads to a larger reduction in emissions and cost.  

To show how the load shifting impacts prices and profits of individual consumers and generators, 
Fig. \ref{fig:price_comparison} plots the profit for generators with the carbon-aware and standard market clearing (left) and the difference in LMPs for different consumers based on results from our equilibrium model and standard model (right). 
In addition, Fig. \ref{fig:RTS-GMLC} illustrates the geographical variation in LMPs at nodes throughout the system and shows where the congested lines are located. 

From Fig. \ref{fig:price_comparison}.a, we observe that most generators experience a reduction in their profits as consumers reduce their demand. This decline in generator profits is more pronounced for carbon-intensive generators than for low-carbon generators.
Further, Fig \ref{fig:price_comparison}.b illustrates that the reduction in demand from carbon-sensitive consumers leads to lower LMPs for most consumers.
This price reduction is primarily driven by the alleviation of transmission congestion, as highlighted in Fig. \ref{fig:RTS-GMLC}. Although the same set of lines (27, 85, 109, and 119) remain congested in both carbon-aware and carbon-agnostic scenarios, congestion levels are lower under carbon-aware conditions. This is reflected in the dual variables of constraint (\ref{eqtra}), which are -43.9, 2.2, 7.9, and 1.3 in the carbon-aware case, compared to -51.4, 2.6, 9.3, and 1.5 in the carbon-agnostic case. Furthermore, Fig. \ref{fig:RTS-GMLC} demonstrates that consumers with higher LMPs (shown by dark purple triangles) experience greater reductions in LMPs (shown by light blue circles) under the carbon-aware scenario than those with lower LMPs (light purple triangles are mostly linked to dark blue circles). 

\begin{figure}[t]
\centering
\subfigure[Generators' profits.]{
\includegraphics[width=0.45\linewidth]{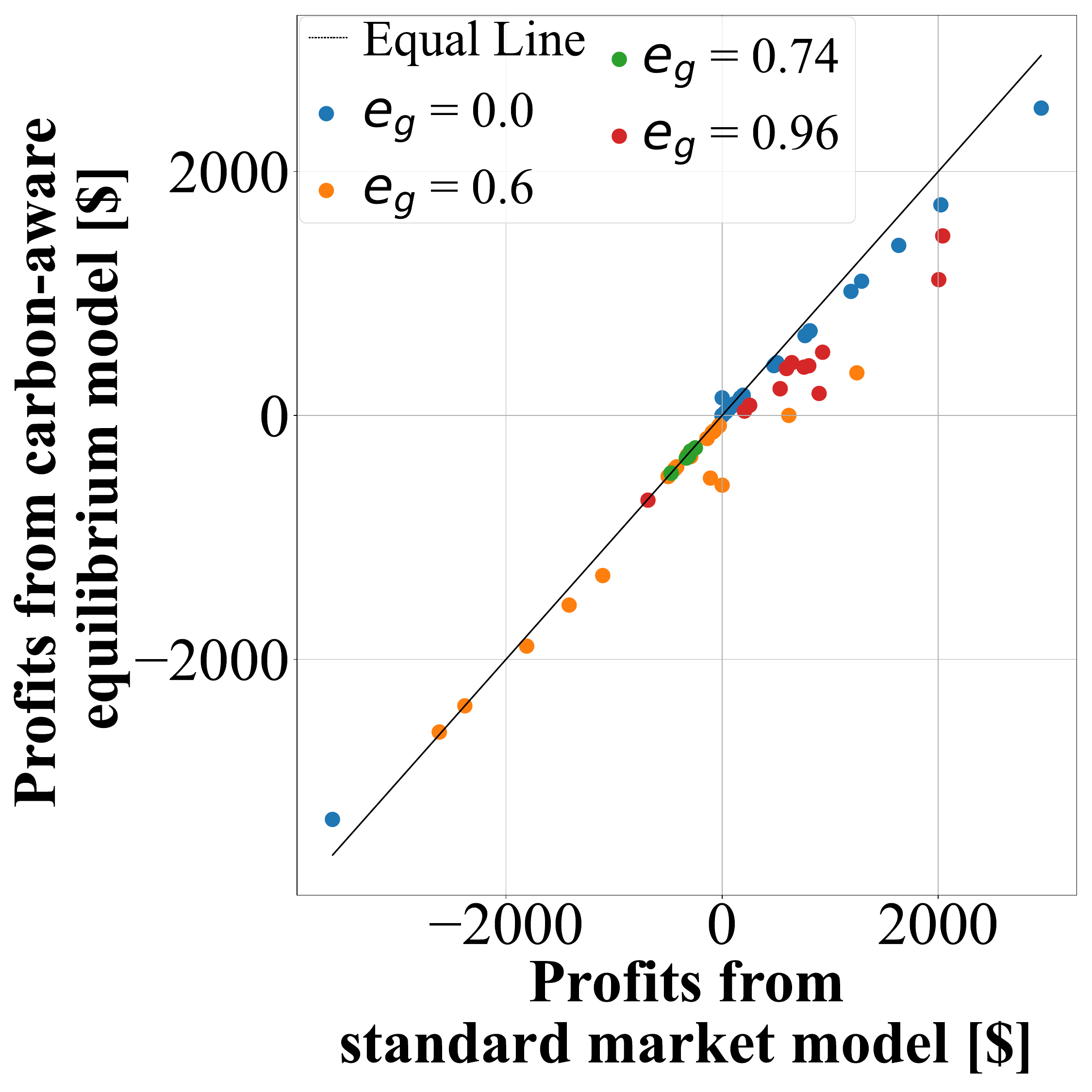}
}
\quad
\subfigure[Consumers' LMPs differences.]{
\includegraphics[width=0.45\linewidth]{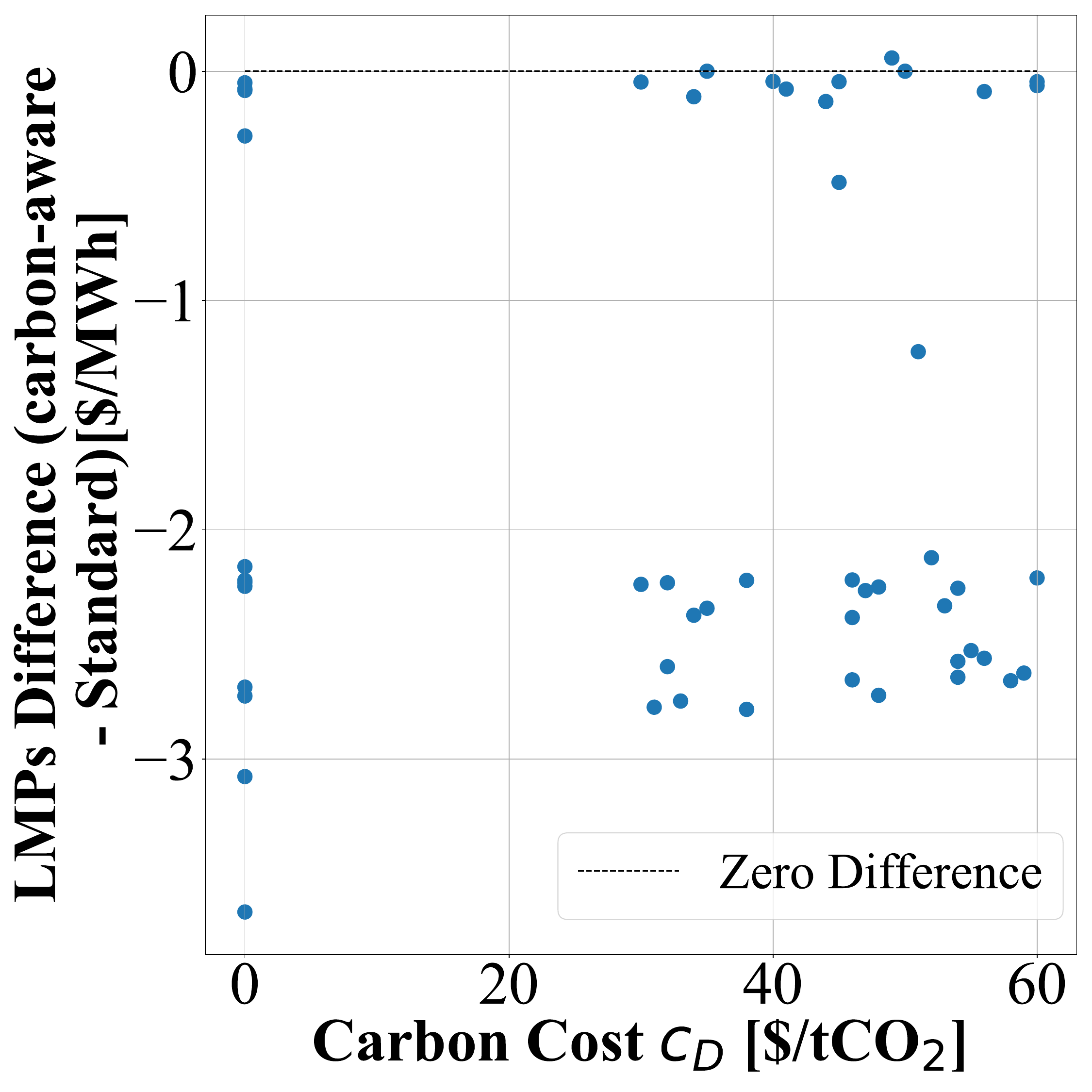}
}

\caption{Impact of carbon costs on generation profits and electricity prices.}
\label{fig:price_comparison}
\end{figure}

\begin{figure}[t]
    \centering
    \includegraphics[width=0.9\linewidth]{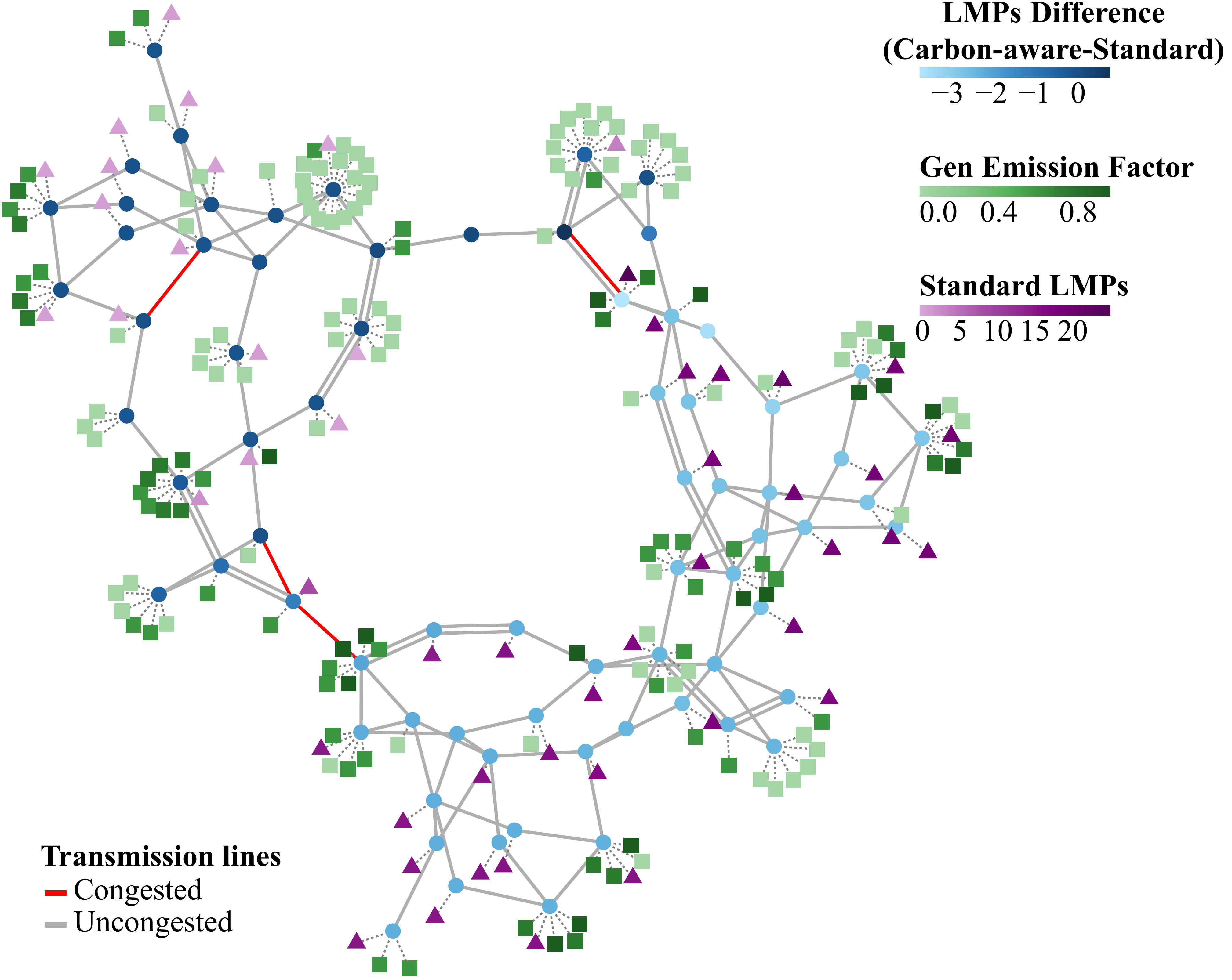}
    \caption{{\blue Changes in LMPs with the introduction of carbon cost. Each circle represents a bus node, with the color intensity representing the difference between carbon-aware and standard LMPs. The squares represent generators, with the color intensity reflecting their carbon emission intensity factors. The triangles attached to each node represent consumers, with the color intensity representing their respective carbon costs. Congested transmission lines are highlighted by the red color (the uncongested lines are in gray).}}
    \label{fig:RTS-GMLC}
\end{figure}

}

\subsection{Comparing against the Sequential Method}
\label{cb}
\begin{figure}[!htpb]
\centering
\subfigure[Total Generation (MWh).]{
\includegraphics[width=\linewidth]{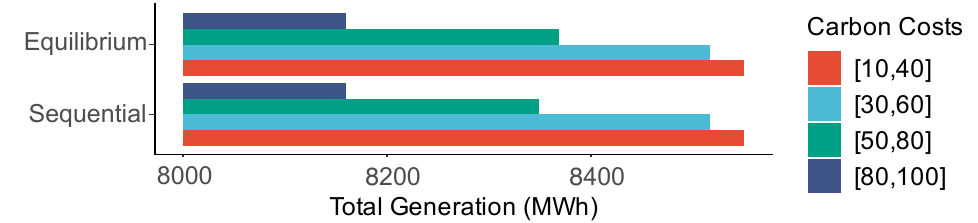}
}

\subfigure[Total Generation Cost (\$).]{
\includegraphics[width=\linewidth]{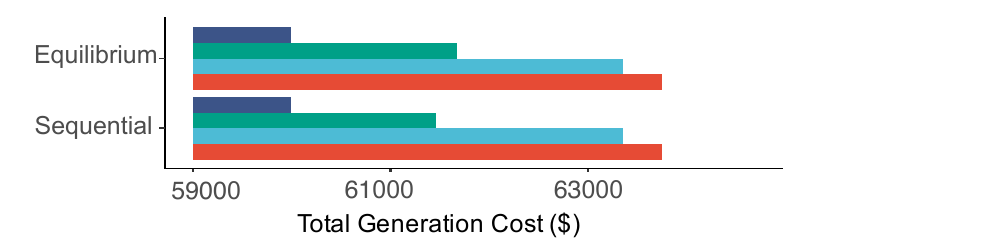}
}

\subfigure[Total Carbon (tCO$_2$).]{
\includegraphics[width=\linewidth]{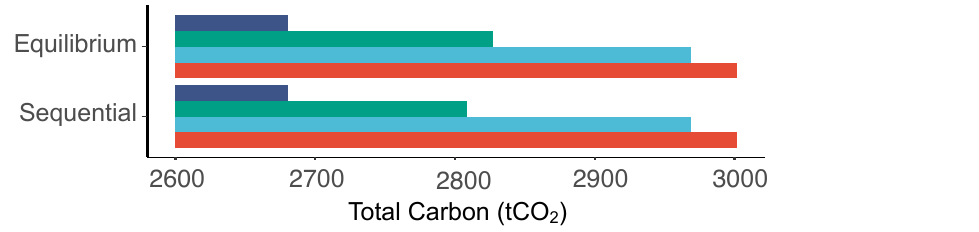}
}

\subfigure[Average Carbon (tCO$_2$/MWh).]{
\includegraphics[width=\linewidth]{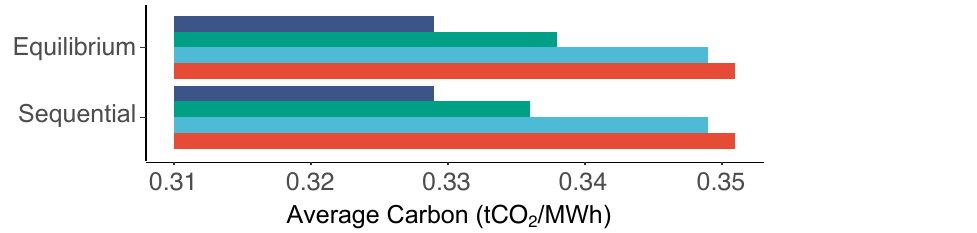}
}

\subfigure[Total Carbon Cost (\$).]{
\includegraphics[width=\linewidth]{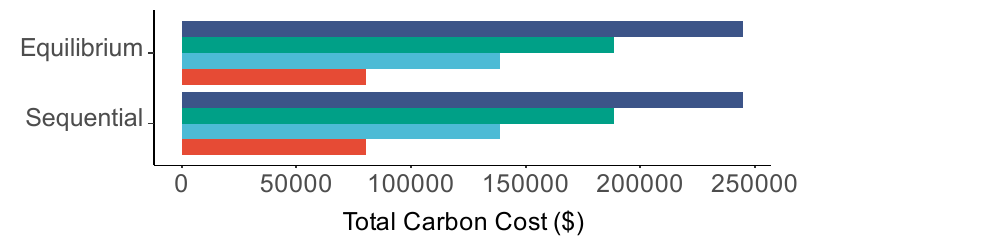}
}
\caption{Comparison of the equilibrium model with the sequential load shifting model.}
\label{figc}
\end{figure}

We next seek to compare our results to those achieved if we use the sequential method. 
We run the sequential method with the same input data as we provided to the equilibrium model in Section \ref{adcb}. The results for both methods under different carbon cost ranges are shown in Fig. \ref{figc}.  


We first focus on the case with carbon costs in $[10,40]$. Due to carbon costs not being high enough, both the equilibrium method and the sequential method share the same optimal solutions compared to carbon-agnostic results. With higher carbon costs in the ranges at or above $[30,60]$, the total generation is reduced due to demand reductions. 
We observe that the results from the equilibrium method and the sequential method are the same for carbon costs of $[30,60]$ and $[80,100]$, but different for the case with $[50,80]$. Specifically, the sequential method has lower total generation (indicating larger demand reduction) and lower average carbon emissions as compared with the equilibrium method. 
We next explain these differences by comparing the average carbon emissions observed by the sequential method when performing load shifting in step $2$ (which is based on the initial market clearing) with the average carbon emissions after the market has cleared again in step $3$, as well as the average carbon emissions from the equilibrium method.


For the [50,80] case, consumers in the sequential model observe an average carbon emission of $\lambda_{b}=0.351$ after the initial market clearing (Step $1$). This $\lambda_{b}$ is high enough for some loads to reduce their consumption from $P_{D,d}^{\max}$ to $P_{D,d}^{\min}$. This leads to a reduction in load, which lowers the average carbon emission to $\lambda_{a}=0.336$ after the market clears again (Step $3$). This $\lambda_{a}$ is low enough for some loads to want to increase their consumption again; however, they do not know this at the time of shifting and are unable to act after the market has cleared the second time (i.e., consumers conduct load shifting in Step 2 only using last clearing results but without knowing the impacts of their behaviors on the next clearing results).
With the equilibrium model, the loads can “see” the impact of their demand reductions when the market is cleared. In this case, the market participants “agree” to a dispatch where the average carbon emissions $\lambda=0.338$. With these average carbon emissions, more loads choose to consume $P_{D,d}^{\max}$ as compared with the sequential load shifting case. This leads to a smaller load reduction, which corresponds to higher total emissions. However, the solution to the equilibrium model is more aligned with the objective of maximizing the (carbon-dependent) utility of the loads. 

Even in the cases where we get similar solutions for the sequential and equilibrium cases, i.e., with carbon cost ranges of $[30,60]$ and $[80,100]$, there are differences between the average carbon emissions $\lambda_{b}$ seen by the loads when doing the shifting (in step 2) and the resulting $\lambda_{a}$ after the market has cleared again, even though the latter is exactly the same as the ones computed by the equilibrium model. 
For example, with carbon costs $[30,60]$, these two $\lambda$ values are $\lambda_{b}=0.351$ while performing the shifting and $\lambda_{a}=0.349$ after the subsequent market clearing. However, this change in $\lambda$ is too small to have a direct impact on the optimal amount of demand reduction. This is because our model exhibits a bang-bang behavior, where loads switch from $P_{D,d}^{\max}$ to $P_{D,d}^{\min}$ once the term $r_{D,d}-p_{i:d\in\mathcal{D}_i}-\lambda\cdot c_{D,d}$ goes from positive to negative. If the change in average carbon emissions is small, it might be that no load experiences this switch. In the case with $[30,60]$, the equilibrium model arrives at an optimal average carbon emission $\lambda=0.349$, and the demand reduction of the equilibrium model matches the demand reduction of the sequential model. This indicates that the change in average carbon emissions from $\lambda_{b}=0.351$ to $\lambda_{a}=0.349$ did not change the optimal consumption level of any load.

\subsection{Analysis for Multi-Period Model}
We then investigate the multi-period model with temporal load shifting and compare results with all benchmark methods. We consider $T=24$ and set the total electricity consumption for each consumer as 
$\mathcal{P}_{D,d} = 24*95\%*P_{D,d}^{\max}$. Parameters $r_{D,d,t}$, $e_{G,g,t}$, and $c_{G,g,t}$ are fixed for each consumer across time. We match the maximum generation of renewable generators with different types of solar PV, solar rooftop PV, and wind with the real-time statistics from \cite{barrows2019ieee} (maximum hydro generation capacities are fixed in our setting). The carbon costs to each consumer are assigned within the range of $[10, 40]$ and fixed across all periods. 
In the sequential method, to make the total electricity consumption for each consumer equal to pre-defined $\mathcal{P}_{D,d}$, we solve a similar MCFD problem by setting the demand of each consumer at $95$\% of their maximum levels across all times (average behavior) in Step 1\footnote{There are also other temporal implementation possibilities for the sequential method. For example, we can also consider cumulative load shifts designed in Section 4.1 of Ref. \cite{lindberg2022using} in a sliding window manner. Basically, this method assumes that the impact of load shifts is applied to the market clearing in the next time step. However, a more detailed discussion of the implementation in the sequential method is out of the scope of this paper.}. We use SM-pre and SM-post to represent SM results after the initial market clearing (Step 1) and SM results after the market clears again (Step 3), respectively.


\begin{figure}[t]
\centering
\subfigure[Average carbon emissions.]{
\includegraphics[width=\linewidth]{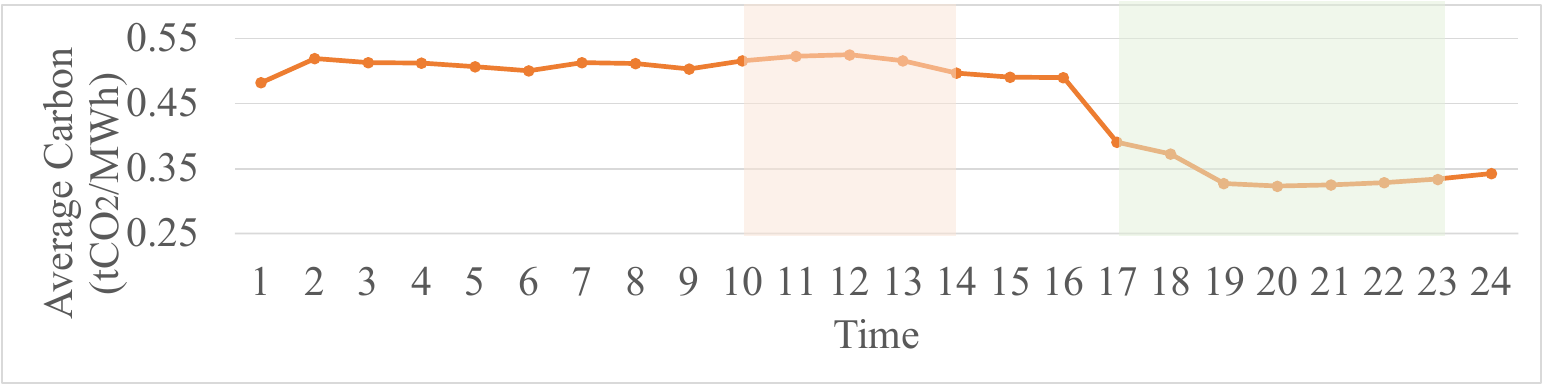}
}


\subfigure[Maximum renewable capacities and renewable generation contributions.]{
\includegraphics[width=\linewidth]{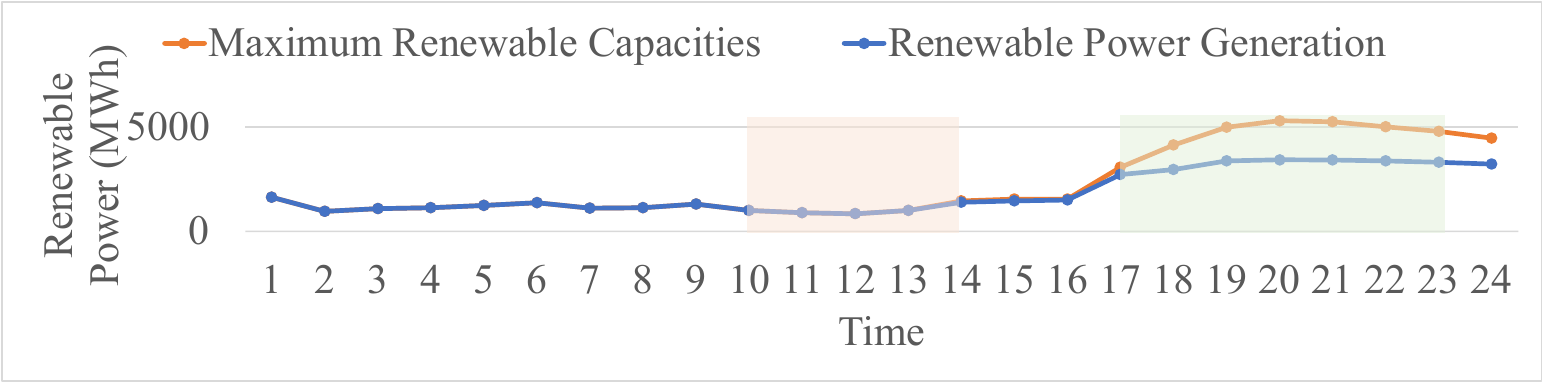}
}

\subfigure[Load dispatch.]{
\includegraphics[width=\linewidth]{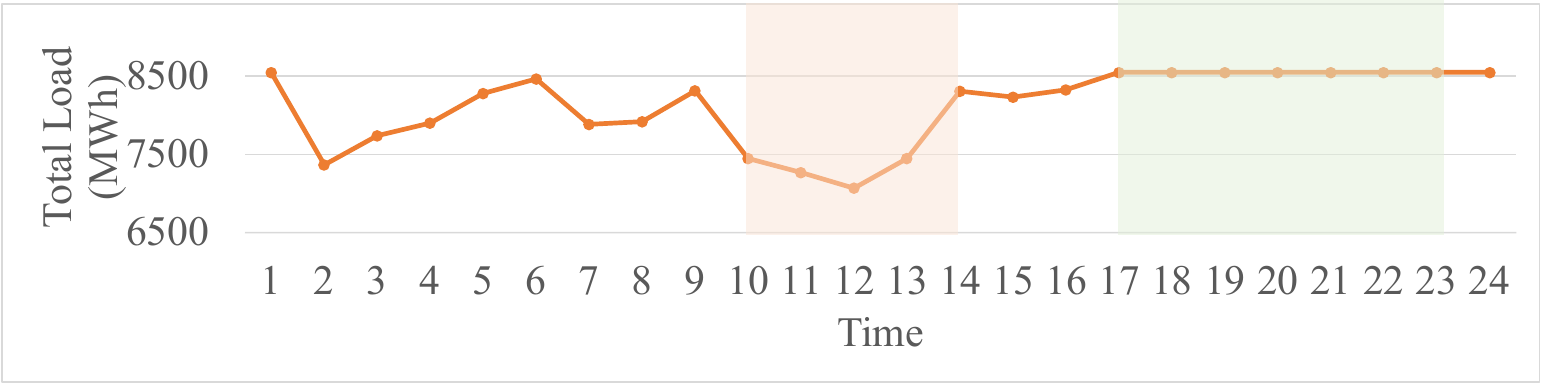}
}


\caption{Average carbon emissions (a), and renewable generation levels (b), and total load dispatch at each time (c). We highlight hours with high average carbon emissions
(orange) and decreasing low average carbon emissions (green).}
\label{figeq}
\end{figure}

\subsubsection{Temporal Load Shifting Results using Our Method} Fig. \ref{figeq} presents average carbon emissions, renewable generation levels, and load dispatch at each time, as computed using our equilibrium method. The maximum renewable capacities in Fig. \ref{figeq} (b) are calculated by summing all maximum available power by solar PV and wind generators for each time. Fig. \ref{figeq} (c) shows the temporal load shifting results across all time by summing all consumers’ demand at each time. Fig. \ref{figeq} highlights hours with high average emissions (orange) and decreasing low average emissions (green).     


Fig. \ref{figeq} (a) and (b) illustrates an inverse relationship between average carbon emissions and renewable generation contributions (or maximum renewable limits). As the average carbon emissions drop from 0.391 tCO$_2$/MWh at $t=17$ to a minimum of 0.323 tCO$_2$/MWh, the renewable power generation increases from 2735MWh to 3426 MWh. Prior to $t=16$, both average carbon emissions and renewable generations fluctuate within relatively narrow bands: carbon emissions between 0.482 tCO$_2$/MWh and 0.525 tCO$_2$/MWh, and renewable generation between approximately 959.5 MWh and 1645.5 MWh. The whole trend of renewable power generation (blue line in b) aligns well with that of the maximum generation capacities (orange line in b). Accordingly, the total consumption in Fig. \ref{figeq} (c) decreases during high-carbon periods (orange) and increases during low-carbon periods (green). 
These results suggest that average carbon emissions are effective to drive consumers to consume more when more renewable generation is available. Note that from the green shadow in Fig. \ref{figeq} (b), when all consumers are operating at their maximum consumption levels, the total amount of renewable power the system can utilize is limited and less than the maximum available renewable generation. 
To further see the carbon reduction performance of our equilibrium method, we will compare our method with benchmarks in the next subsection.

\begin{table}[t]\footnotesize
\centering
\renewcommand{\arraystretch}{1}
\setlength{\tabcolsep}{1.6pt}
\caption{Average results for temporal load shifting based on different methods.}
\label{tab41}
\begin{tabular}{cccccc}
\hline
Methods& \begin{tabular}[c]{@{}c@{}}Total Generation\\~[MWh]\end{tabular}  &\begin{tabular}[c]{@{}c@{}}Total Generation \\ Cost [\$]\end{tabular} &\begin{tabular}[c]{@{}c@{}}Total Carbon\\~[tCO$_2$]\end{tabular}&\begin{tabular}[c]{@{}c@{}}Average Carbon\\~[tCO$_2$/MWh]\end{tabular}\\
\hline
SM-pre	&	&1503033.0	&71526.5	&0.3669\\
SM-post	&	&1469790.0	&69266.7	&0.3553\\
CTM &&1493842.5&65733.8	&0.3372\\
MCE &	&1536474.1 &64750.4 &0.3322\\
Ours&	\multicolumn{1}{c}{\multirow{-5}{*}{194940}}	&1463917.1	&69066.9	&0.3543	\\
\hline
\end{tabular}
\end{table}

\subsubsection{Comparison with Benchmark Methods} 
\begin{figure}[t]
\centering
\subfigure[Total generation cost.]{
\includegraphics[width=\linewidth]{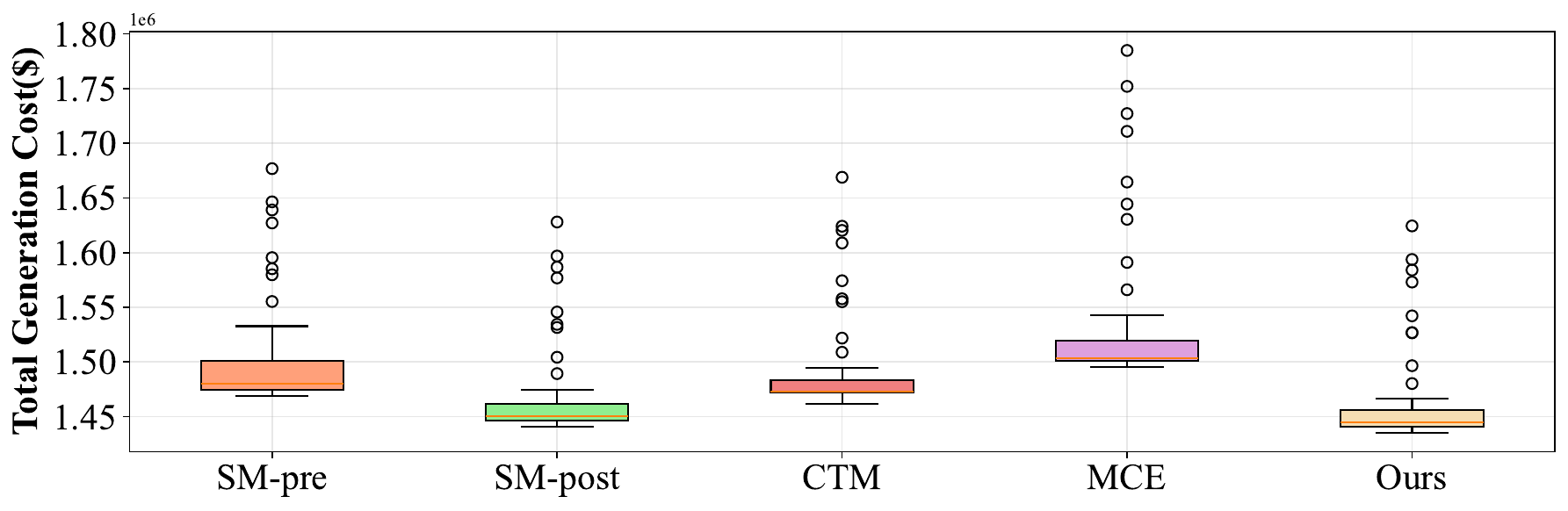}
}


\subfigure[Total carbon emissions.]{
\includegraphics[width=\linewidth]{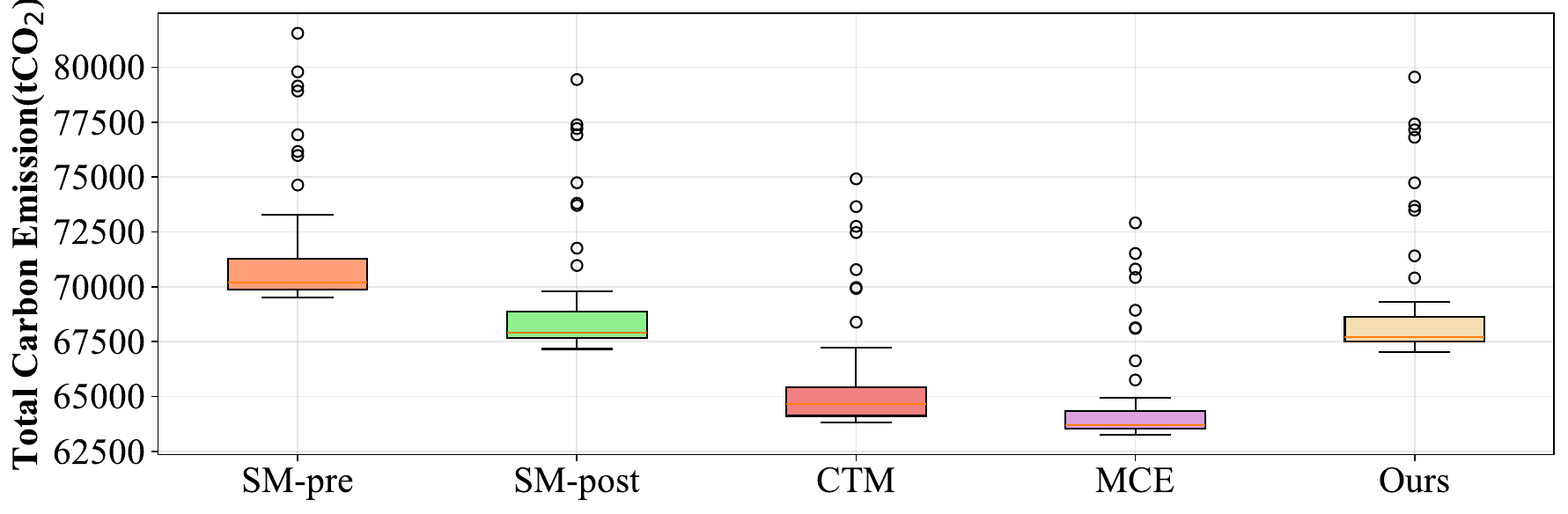}
}

\subfigure[Temporal load shifting (90\% confidence level).]{
\includegraphics[width=\linewidth]{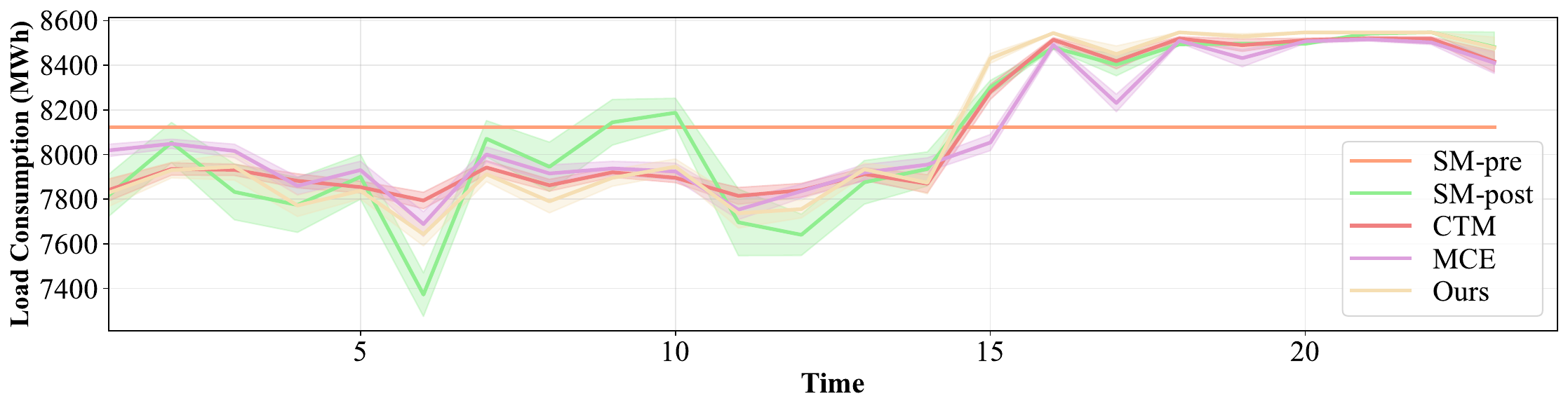}
}
\caption{Results for temporal load shifting based on different methods.}
\label{figeq2}
\end{figure}

We set the carbon tax as $C_{co_2}=20$\$/MWh. We test 50 scenarios with different carbon cost combinations in the range of $[0,40]$ and maximum renewable capacities dynamics. Table \ref{tab41} shows average results for all scenarios. Fig. \ref{figeq2} (a) and (b) show more detailed distribution results for total generation cost and total carbon emissions (or average carbon emissions because the total generation is the same across different scenarios with 194,940 MWh), respectively. Fig. \ref{figeq2} (c) shows temporal load shifting results, where the solid line represents the mean, and the shadow area represents the 90\% confidence interval. 


As shown in Table \ref{tab41}, our method achieves the lowest total generation cost of \$1,463,917.1, outperforming all benchmark methods, corresponding to cost reductions of 2.6\% compared to SM-pre, 0.4\% compared to SM-post, 2\% compared to CTM, and 4.7\% compared to MCE. From Fig. \ref{figeq2}(a), our method exhibits sensitivity to carbon costs and renewable dynamics comparable to SM-post, greater than that of CTM, but less than that of SM-pre and MCE. 
In terms of carbon emissions, our method yields a total of 69,066.9 tCO$_2$, corresponding to carbon reductions of 3.4\% compared to SM-pre, 0.3\% compared to SM-post, and carbon increases of 5.1\% compared to CTM, 6.7\% compared to MCE. 
From Fig. \ref{figeq2}(b), our method exhibits lower sensitivity to carbon costs and renewable dynamics than both SM (SM-pre and SM-post) and CTM, but higher sensitivity compared to MCE.
Although CTM achieves a lower absolute emission value at 65,733.8 tCO$_2$, it does so at a higher generation cost compared to our method. From Fig. \ref{figeq2}(c), after $t=17$, when more renewable power is available, the load shifting behavior of our method closely aligns with that of CTM and SM-post. However, notable differences emerge before t=17, when available renewable power is less. CTM reduces carbon emissions by impacting generation merit order, while both our method and SM reduce carbon emissions by shifting loads. Although impact mechanisms are different, they drive the similar consumption patterns when more renewable power is available. 
However, when compared with MCE, minimizing the total carbon emission does not necessarily mean all consumers should consume at their maximum levels when there is more available renewable generation (different load shifting results after $t=17$ compared to the previous three methods). Fig. \ref{figeq2}(c) also reveals that, relative to the average behavior from SM-pre, both our method and SM-post shift more consumption to the high-renewable period after $t=17$. However, since load shifting in SM relies on average carbon emissions computed based on market clearing results from the last step without knowing the current carbon information influenced by other consumers, its consumption pattern exhibits higher fluctuation than ours before $t=17$. 

\section{Conclusions}
\label{sec6}


{\blue This paper designs a novel carbon-aware equilibrium model to analyze the impact of load shifting guided by carbon signals on emission reductions. Our case study reveals that guiding load shifting with an average carbon signal can be ineffective, or even counterproductive, for reducing emissions. This is because it fails to reflect marginal carbon impacts, ignores geographical differences, and may create new emissions-intensive demand peaks. 
Further, our results demonstrate that the equilibrium method is equivalent to the sequential method when carbon costs of consumers are in high or low ranges.

} 




{\blue This paper outlines several promising directions for future research. First, given the observed limitations of average carbon emission signals in driving effective decarbonization, we plan to extend our analysis to explore alternative signals. These include locational carbon emission signals \cite{sofia2024carbon, chen2024contributions, lindberg2020environmental, chen2024carbon, jiang2025greening} and other improved carbon emission signals aimed at enhancing emission reduction outcomes. Second, while our equilibrium model captures the behavior of carbon-sensitive consumers who maximize their carbon-dependent utility in response to average carbon emission signals, it remains unclear whether the actions of these consumers also maximize the carbon-aware system-level social welfare. 
So far, we have not been able to find an equivalent single optimization problem for our equilibrium problem, and such an equivalent problem may not exist, making it more challenging to analyze whether the equilibrium problem maximizes some definition of carbon-aware social welfare.
Furthermore, to the best of our knowledge, a well-defined and widely accepted definition of carbon-aware social welfare is still lacking in the literature. 
A valuable and interesting direction for future work would be to define such an objective and investigate how carbon emission signals can be designed to align consumers' carbon-aware behavior with the system-level carbon-aware social welfare. Finally, in the current formulation, the carbon cost is modeled as an exogenous parameter submitted and randomly chosen by each consumer (e.g., uniformly distributed in $[0,100]$ for numerical studies). Since carbon cost values critically impact market clearing results, leading to varying electricity bills and carbon emissions, it is worth further investigating how consumers strategically set carbon costs to balance their individual economic and environmental goals and what kinds of carbon cost combinations are preferred from a system-level perspective.}

\bibliographystyle{ACM-Reference-Format}
\appendix

\section{Alternative Equilibrium Formulation for Single-Period Model}
\label{appenef}
The single-period equilibrium problem, characterized by parametric optimization problems \eqref{eq22}, \eqref{eq33}, \eqref{eqtr} and the parametric complementarity conditions \eqref{eq77c} and \eqref{eq923}, can also be represented in an alternative way that more clearly demonstrates the connection to standard, OPF-based electricity market clearing. 
If we combine the Karush-Kuhn–Tucker (KKT) optimality conditions for problems \eqref{eq22}, \eqref{eqtr}, and the condition \eqref{eq77c} in one problem, we observe that these three roblems combined are equivalent to the standard DCOPF problem, i.e., 
\begin{subequations}
\label{eqaneq}
    \begin{align}
    \min_{P_G, \theta} \ &\sum_{g\in \mathcal{G}}c_{G,g}P_{G,g}\\\
    s.t. \ & \sum_{d\in \mathcal{D}_i}P_{D,d}+\sum_{\mathclap{j:(i,j)\in \mathcal{L}}}\beta_{ij}(\theta_i-\theta_j)=\sum_{g\in \mathcal{G}_i}P_{G,g},\ \forall i \in \mathcal{N},\\
    &P_{G,g}^{\min}\le P_{G,g}\le P_{G,g}^{\max},\ \forall g\in \mathcal{G},\\
    &-F_{ij}^{\rm{lim}}\leq\beta_{ij}(\theta_i-\theta_j)\leq F_{ij}^{\rm{lim}}, \quad\forall (i,j)\in \mathcal{L},\\
    &\theta_{ref} = 0.
\end{align}
\end{subequations}
This shows that an equilibrium problem involving the standard DCOPF problem \eqref{eqaneq}, the load shifting problem \eqref{eq33}, and the parametric complementarity condition \eqref{eq923} that defines the average carbon emissions is equivalent to the equilibruim problem defined by optimization problems \eqref{eq22}, \eqref{eq33}, \eqref{eqtr} and the parametric complementarity conditions \eqref{eq77c} and \eqref{eq923}, and thus, have the same optimal solutions.

This alternative formulation is helpful to understand the relationship between the sequential method and the equilibrium method. The sequential method (as outlined in Section \ref{sec41}) solves the OPF problem and then calculates the average carbon emissions, which subsequently are used as an input for consumers' load dispatch. This implies that consumers are shifting their loads without knowing how their actions will impact the average carbon values. In contrast, the equilibrium method solves the OPF problem, consumers' problem, and average carbon emissions condition simultaneously, representing an ideal scenario where consumers know the true post-shift average carbon value.

\section{Equilibrium Formulation for Carbon Tax Method}
\label{appen2}
In the equilibrium formulation of the carbon tax method, there are four participants: generators, consumers, transmission owners, and ISO. Transmission owners maximize their profit by solving the same problem (\ref{eqtr}) in our method and ISO clears the market by determining the electricity price with the parametric complementarity condition (\ref{eq77c}). Different from our method, each generator in the carbon method aims to maximize profit under both price and carbon emissions, while each consumer only considers price information in their decision-making. These two problems are listed as follows.

\textbf{Generators:}
\begin{subequations}
\label{eq22-c}
    \begin{align}
    \max_{P_{G,g}} \ &(p_{i:g\in \mathcal{G}_i}-c_{G,g}-C_{co_2}e_{G,g})\cdot P_{G,g}\label{eq22obja}\\
    s.t. \ & \rm{Constraint} \ (\ref{eq22a}).\notag
\end{align}
\end{subequations}

\textbf{Consumers:}\begin{subequations}
\label{eq33-c}
\begin{align}
    \max_{P_{D,d}} \ &(r_{D,d}-p_{i:d\in \mathcal{D}_i})\cdot P_{D,d}\label{eq33obja}\\
    s.t. \ &\rm{Constraint} \ (\ref{eq33a}).\notag
\end{align}
\end{subequations}

Compared to our equilibrium method, where consumers make their decision-making considering carbon emissions, the carbon tax method adds the carbon consideration into generators' dispatch. As generators roll carbon taxes into their generation costs, the carbon tax method increases the cost of carbon-heavy generation sources and provides a competitive advantage to cleaner generators, such as solar and wind power generators. Since some generators emit more carbon than others for unit power generation, adding a carbon tax on generation might change the merit order of the generators and promote lower carbon generation dispatch solutions. In contrast, our equilibrium method tries to shift consumers' demand to lower-carbon generation time, promoting lower carbon power consumption. 

{\blue \section{Existence and Uniqueness of Solutions}
\label{appendiex} 
From the Nash equilibrium problem, which includes the parametric convex optimization problems~\eqref{eq22}, \eqref{eq33}, \eqref{eqtr} and the parametric complementarity conditions~\eqref{eq77c} and~\eqref{eq923}, an equivalent variational inequality (VI) problem denoted by $VI(\mathcal{X},F)$ can be derived by concatenating the optimality conditions of the optimization problems and the complementarity conditions. That is, the VI problem consists of finding a vector $x^*\in \mathcal{X}$ such that
\begin{equation}
    F(x^*)^ \intercal(x-x^*)\ge 0,\ \forall x\in \mathcal{X},
\end{equation}
where $x = (P_{G,g}, \forall g\in \mathcal{G}; P_{D,d}, \forall d\in \mathcal{D}; \theta_i, \forall i \in \mathcal{N}; p_i,\forall i \in \mathcal{N}; \lambda)$, $\mathcal{X} = [\text{ Constraints } (1b),(2b) \text{ and } (3b);\mathbb{R}^{|\mathcal{N}|};\mathbb{R}]$, and 
\[
F(\mathbf{x}) =
\begin{pmatrix}
[c_{G,g}-p_{i:g\in \mathcal{G}_i}]_g \\
[\lambda c_{D,d}+p_{i:d\in \mathcal{D}_i}-r_{D,d}]_d \\
[p_i\sum_{j:(i,j)\in \mathcal{L}}\beta_{ij}-\sum_{j:(i,j)\in \mathcal{L}}p_j\beta_{ij}]_i\\
[\sum_{d\in \mathcal{D}_i}P_{D,d}+\sum_{j:(i,j)\in \mathcal{L}}\beta_{ij}(\theta_i-\theta_j)-\sum_{g\in \mathcal{G}_i}P_{G,g}]_i\\
\lambda\sum_{d \in \mathcal{D}} P_{D,d} - \sum_{g \in \mathcal{G}}e_{G,g}\cdot P_{G,g}
\end{pmatrix}.
\]

\subsection{Solution Existence}
A basic result on the existence of a solution to the variational inequality problem $VI(\mathcal{X}, F)$ is as follows.

\textbf{Theorem 1}\cite{eaves1971basic,hartman1966some}: Let $\mathcal{X}$ be a nonempty, compact, and convex subset of $\mathbb{R}^n$ and
let $F$ be a continuous mapping from $\mathcal{X}$ to $\mathbb{R}^n$. Then, there exists a solution to the problem $VI(\mathcal{X}, F)$.

{\blue
Our VI fulfills the hypothesis of Theorem 1, save for the compactness of $\mathcal{X}$.

Indeed, Constraints~\eqref{eq22a},~\eqref{eq33a}, and~\eqref{eqtra} ensure the boundedness of variables $P_{G,g}$, $P_{D,d}$, and $\theta_i$, respectively, in $\mathcal{X}$.
However, this is not the case for either $p$ or $\lambda$.
We shall pursue the classical strategy to show that any solution of our Nash equilibrium problem is uniformly bounded in norm by $R_0$.
This implies that the $VI(\mathcal{X}, F)$ and $VI(\mathcal{X}\cap\mathbb{B}_{R}, F)$,
where $\mathbb{B}_R$ is the ball of radius $R$, have the same solution set for all $R\geq R_0$.
Then, Theorem~1 is applicable to the latter VI, and then yields the existence result for the Nash equilibrium.

We start by showing that for any $x\in\mathcal{X}$, its last component, corresponding to $\lambda$, is bounded.
First, it is clear that $\lambda \geq 0$ from~\eqref{eq923}. For the upper bound, we note that
\begin{align*}
   \lambda &= \frac{\sum_{g\in \mathcal{G}}e_{G,g}P_{G,g}}{\sum_{d\in \mathcal{D}}P_{D,d}}\le e^{\max}_{G} \frac{\sum_{g\in \mathcal{G}} P_{G,g}}{\sum_{d\in \mathcal{D}}P_{D,d}}
   \intertext{where $e^{\max}_{G}$ is the maximum value of any $e_{G,g}$.
   Since~\eqref{eq77c} holds for all $i$, summing these equality yields $\sum_{g\in \mathcal{G}}P_{G,g} = \sum_{d\in \mathcal{D}}P_{D,d}$, which gives}
   \lambda &\le e^{\max}_{G}.
\end{align*}
Turning our attention to the prices $p$, observe that fixing $\lambda$ in our Nash equilibrium enables us to use the classical welfare theorem.
This enables us to formulate the following necessary condition on any given $x\in\mathcal{X}$ to be a solution: 
the powers $\{P_{G,g}\}$, $\{P_{D,d}\}$, voltage angles $\theta$ and prices $p$ must be a solution to the welfare maximization problem
\begin{subequations}
\label{opt:social_welfare}
   \begin{align}
    \max_{P_{G},P_{D},\theta} & \sum_{d\in \mathcal{D}}(r_{D,d}-c_{D,d}\lambda) P_{D,d} - \sum_{g\in \mathcal{G}} c_{G,g}P_{G,g}\label{obj:social_welfare}\\
    & \text{Constraints}~\eqref{eq22a},~\eqref{eq33a},~\eqref{eqtra},~\eqref{eq77c} ~\text{ hold.}
\end{align} 
\end{subequations}

We now show that the norm of $p$ is uniformly bounded for any $\lambda\in[0,e^{\max}_{G}]$.
For this purpose, we compactly write down the Problem~\eqref{opt:social_welfare} in the form
\begin{equation}\label{opt:max_duality}
   \max_{z} \qquad \langle \cL, z\rangle\qquad\text{such that}\qquad z\in\boxZ \qquad \A z = 0,
\end{equation}
where $z$ collects all powers $P_{D,d}$, $P_{G,g}$, and voltage angle $\theta_i$.
The cost vector $\cL$ collects the terms in the objective function of~\eqref{obj:social_welfare},
while $\A$ (resp. $\boxZ$) captures the network constraint~\eqref{eq77c} (resp. simple bounds constraints~\eqref{eq22a},~\eqref{eq33a},~\eqref{eqtra}).
Let $\zOP$ denote the operating point satisfying~\eqref{eq77c} such that no variable of the above problem is at a bound whose existence we assumed in Section \ref{sub-21}.
In the current notation, this assumption corresponds to the existence of $\zOP$ such that $\A\zOP = 0$ and $\zOP\in\INT\boxZ$,
where $\INT$ denote the interior of a set.

The proof has the following parts:
\begin{enumerate}
    \item We show that the matrix $A$ that appears in the equality constraints of Problem~\eqref{opt:max_duality} is surjective.
    \item Using the Fenchel duality scheme, we explicitly derive the dual problem to~\eqref{opt:max_duality}.
    \item We derive several inequalities as a preliminary step to establishing a uniform estimate
    on the price norm $||p||$ for any $\lambda\in [0,e^{\max}_G]$. 
    \item We derive the uniform estimate on the price norm $||p||$.
    \item We conclude on the existence of the aforementioned $R_0$ that any solution to
    the Nash equilibrium problem must satisfy. This allows us to use Theorem~1 to show the existence of a solution to our developed Nash equilibrium problem.
\end{enumerate}
\textbf{Part 1} Let us show that $\A$ is surjective (or onto): the equation~\eqref{eq77c} can be written in a vector form as
$[E_{D} \quad -E_{G} \quad L_w^\mathrm{red}]z = 0$, where $ L_w^\mathrm{red}$ is the weighted Laplacian (bus admittance metrix in power system)
with the column corresponding to $\theta_{\mathrm{ref}}$ removed, and each column of both $E_{D}$ and $E_{G}$ is
a unit vector. Then, since the network is supposed to be connected, the kernel of $(L_w^\mathrm{red})^T$
is spanned by $e$, the vector of ones. Since the dot product of $e$ and any unit vector is nonzero,
then $\A^T$ has a trivial kernel and $\A$ is surjective.


\textbf{Part 2} To investigate the qualitative behavior of the norm of the prices $p$, we use the Fenchel duality scheme, as laid out in
\cite[Example~11.41, p.~505]{rockafellar2009variational}.
For this, let us introduce the following notations: for any set $S$, let $\INDF{S}$ be the indicator function of $S$,
defined as $\INDF{S}(u)=0$ whenever $u\in S$ and $+\infty$ otherwise.
Its Fenchel conjugate is $\SFF{S}(u)=\sup_{v\in S}\langle v,u\rangle$,
the support function of the set $S$. 
Applying the Fenchel duality scheme to the Problem~\eqref{opt:max_duality}, we have that
\begin{align}
   \max_z\; &-\langle-\cL,z\rangle - \INDF{\boxZ}(z) - \INDF{\{0\}}(Az)\label{opt:primal_welfare}
   \intertext{and}
   \min_p\; &\SFF{\boxZ}(-\cL -A^T p)\label{opt:dual_price}
   \intertext{have the same objective function value and their solution sets are bounded whenever}
   0 &\in \INT (\A^T\mathbb{R}^{|\mathcal{N}|} + \dom \SFF{\boxZ})\label{int1}\\
   -\cL&\in\INT(\A \boxZ )\label{int2},
\end{align}
where $\dom$ denotes the \emph{effective domain} in the sense of convex analysis.
Note that the interiority condition~\eqref{int1} is trivially satisfied as $\boxZ$ is a bounded set,
which implies that $\SFF{\boxZ}$ has full effective domain
(that is $\SFF{Z}(u)$ is finite for all $u$).
The second one follows the existence of $\zOP$ and the surjectivity of $\A$.
This yields that the prices are bounded for a fixed $\lambda$.

\textbf{Part3} However, we need a uniform bound on $\|p\|$ w.r.t. $\lambda\in[0,e^{\max}_G]$.
To show this, we use the following classical approach:
We leverage the coercivity (in the extended-real valued sense) of the objective function of the dual problem~\eqref{opt:dual_price}.
This allows us to come up with explicit upper bounds that hold for all values of $\lambda$ of interest.
Let us start with a few preliminary observations: The existence of $\zOP$ enables us to write
$\boxZ=\{\zOP\} + \boxZred$, namely that $\boxZ$ is the Minkowski sum of $\zOP$ and $\boxZred$,
a bounded set with non-empty interior. This leads to the existence of $\varepsilon > 0$ such
that $\varepsilon\mathbb{B}\subseteq\boxZred$, where $\mathbb{B}$ is the unit ball.
A basic property of support function yields that
\begin{equation}\label{ineq:sf}
\SFF{\varepsilon\mathbb{B}}(u)\leq \SFF{\boxZred}(u)\qquad\forall u.
\end{equation}
Moreover, we have the relation
\begin{align}
   \SFF{\boxZ}(-\At p -c) &= \sup_{z\in\boxZ}\;\langle z, -\At p - c\rangle\\
&= \langle\zOP, -\At p - c\rangle + \sup_{z\in\boxZred}\;\langle z, -\At p - c\rangle\\
&= \langle\zOP,- c\rangle + \sup_{z\in\boxZred}\;\langle z, -\At p - c\rangle,
\intertext{where we used the fact that $A\zOP = 0$ since $\zOP$ satisfies the network constraints.
Furthermore, we can rewrite the previous identity as}
\SFF{\boxZ}(-\At p -c)&= \langle\zOP,- c\rangle + \SFF{\boxZred}(-\At p - c).
\intertext{Using the inequality~\eqref{ineq:sf}, we get}
&\geq \langle\zOP,- c\rangle + \SFF{\boxZred}(-\At p - c),
\intertext{which can be equivalently written as}
&\geq \langle\zOP,- c\rangle + \varepsilon\|-\At p - c\|.
\intertext{Finally, using the reverse triangle inequality, we get that}
\SFF{\boxZ}(-\At p -c)&\geq \langle\zOP,- c\rangle +  \varepsilon (\|\At p\| - \|c\|).\label{ineq:SF}
\end{align}

\textbf{Part 4} We are now ready to establish the upper bound on $\|p\|$.
To streamline the proof, let us directly show that any solution to~\eqref{opt:dual_price} satisfies
\begin{equation}\label{est:Atp}
   \|\At p\| < (1+\varepsilon^{-1}2\boxZdiam)\cLmax,
\end{equation}
where $\boxZdiam$ is the maximum norm of any element of $\boxZ$ and $\cLmax = \max_{\lambda\in[0,e_G^{\max}]}\|\cL\|$.

We proceed by contradiction: given $\Delta > 0$ and $\lambda\in[0,e_G^{\max}]$, suppose that $\thep$ is a solution to~\eqref{opt:dual_price} and satisfies
\begin{equation}
   \|\At\thep\| \geq (1+\varepsilon^{-1}2\boxZdiam)\|\cL\| + \varepsilon^{-1}\Delta.
\end{equation}
With $\thez$ a solution to~\eqref{opt:primal_welfare}, note that $2\boxZdiam\|\cL\|\geq\langle \thez + \zOP, \cL\rangle$
by the Cauchy--Schwarz and the triangle inequality. Then,
\begin{gather}
   \|\At\thep\| \geq \varepsilon^{-1}(\langle \zOP + \thez , \cL\rangle + \Delta) + \|\cL\|.
   \intertext{Rearranging terms, we get}
   \varepsilon (\|\At\thep\| - \|\cL\|) + \langle \zOP, -\cL\rangle \geq \langle \thez, \cL\rangle + \Delta.
   \intertext{Using~\eqref{ineq:SF}, we get}
\SFF{\boxZ}(-\At\thep-\cL) \geq \langle \thez, \cL\rangle + \Delta.
\end{gather}
Since $\langle \thez, \cL\rangle$ is the optimal value of~\eqref{opt:dual_price},
we reach the contradiction.

To finish our proof, we need to show that $\|\thep\|$ is uniformly bounded (w.r.t $\cLmax$).
Since $A^T$ has a trivial kernel, the map $x\mapsto\|A^Tx\|$ is a norm.
The equivalence of norms in finite dimension implies that the upper bound~\eqref{est:Atp}
yields a similar upper bound on $\|\thep\|$.

\textbf{Part 5} Coming back to the existence proof of the VI, 
this guarantees that the prices at a solution to the VI are uniformly bounded.
This implies the existence of an $R_0$, a strict upper bound on the size of the solution set of the VI.
Applying Theorem~1 ensures the existence of a solution to the $VI(\mathcal{X}\cap\mathbb{B}_{R_0}, F)$,
and therefore to the Nash equilibrium problem.


\subsection{Solution Uniqueness}
Turning our attention to the uniqueness question, the situation is more delicate.
General uniqueness results for VI usually require the VI mapping to be strictly monotone
over $\mathcal{X}$. However, this property is not enjoyed by the mapping $F$
we consider here. To see this, take any $\hat{x}$, $\tilde{x}\in\mathcal{X}$ where
the prices and $\lambda$ components are identical,
and the other ones are in the neighborhood of $\zOP$.
Then it is easy to check that $(\hat{x}-\tilde{x})^T (F(\hat{x}) - F(\tilde{x})) = 0$,
and $F$ is not strictly monotone on $\mathcal{X}$.

}

}

\end{document}